\newif\ifTwoColumn%
\newif\ifSUBMIT%
\newif\ifCOMMENTS%
\newif\ifFIGs%
\newif\ifFIGoneColumn%
\let\ifTwoColumn\iftrue%
\let\ifSUBMIT\iftrue%
\let\ifCOMMENTS\iffalse%
\let\ifFIGs\iftrue%
\let\ifFIGoneColumn\iftrue%
\ifTwoColumn%
  \documentclass[aps,showpacs,pra,showkeys,twocolumn,superscriptaddress,
     float,floatfix]{revtex4-1} 
\else
  \documentclass{revtex-4.1}
\fi
\usepackage{graphicx}
\usepackage{amsmath}
\usepackage{xcolor}
\usepackage{hyperref}
\usepackage[utf8]{inputenc}
\usepackage{todonotes}
\usepackage{enumitem}

\usepackage{array,mathtools,amssymb,booktabs}

\ifSUBMIT%
  \ifCOMMENTS%
    \usepackage{color}    
    \usepackage[normalem]{ulem}
    \def\EDITS#1{{\color{red}#1}}
    \def\STRIKE#1{{\color{red}\sout{#1}}}
    \def\NEDITS#1{{\color{blue}#1}}
    \def\NSTRIKE#1{{\color{blue}\sout{#1}}}
  \else
    \def\EDITS#1{#1}
    \def\STRIKE#1{}
    \def\NEDITS#1{#1}
    \def\NSTRIKE#1{}
  \fi
\else
  \usepackage{color}    
  \usepackage[normalem]{ulem}
 \definecolor{mygreen}{RGB}{0,180,0}    
  \def\EDITS#1{{\color{red}#1}}
  \def\STRIKE#1{{\color{red}\sout{#1}}}
  \def\NEDITS#1{{\color{blue}#1}}
  \def\NSTRIKE#1{{\color{blue}\sout{#1}}}
\fi

\definecolor{mygray}{RGB}{128,128,128}

\begin{document}

\title{Binary contraction method for the construction of time-dependent
dividing surfaces in driven chemical reactions}
\author{Robin Bardakcioglu}
\author{Andrej Junginger}
\altaffiliation[Present address: ]{Machine Learning Team at ETAS GmbH, Bosch Group}
\author{Matthias Feldmaier}
\author{J\"org Main}
\affiliation{%
Institut f\"ur Theoretische Physik 1,
Universit\"at Stuttgart,
70550 Stuttgart,
Germany}
\author{Rigoberto Hernandez}
\email[Correspondence to: ]{r.hernandez@jhu.edu}
\affiliation{%
Department of Chemistry,
Johns Hopkins University,
Baltimore, Maryland 21218, USA
}
\date{\today}

\newcommand{\EQ}{Eq.}
\newcommand{\EQS}{Eqs.}
\newcommand{\FIG}{Fig.}
\newcommand{\FIGS}{Figs.}
\newcommand{\REF}{Ref.}
\newcommand{\REFS}{Refs.}
\newcommand{\SEC}{Sec.}
\newcommand{\SECS}{Secs.}
\newcommand{\eg}{e.\,g.}
\newcommand{\cf}{cf.}
\newcommand{\ie}{i.\,e.}
\newcommand{\ud}{\mathrm{d}}
\newcommand{\ue}{\mathrm{e}}
\newcommand{\kB}{k_\mathrm{B}}
\newcommand{\VLiCN}{V_\mathrm{LiCN}}
\newcommand{\VCN}{V_\mathrm{C-N}}
\newcommand{\VLi}{V_\mathrm{Li-CN}}
\renewcommand{\vec}[1]{\boldsymbol{#1}}
\newcommand{\qq}{\vec{q}}
\newcommand{\xx}{\vec{x}}
\newcommand{\vv}{\vec{v}}
\newcommand{\transpose}{\mathsf{T}}
\newcommand{\reactantpop}{\mathcal{P}}
\newcommand{\kf}{k_\mathrm{f}}
\newcommand{\etal}{\emph{et al.}}
\newcommand{\LD}{\mathcal{L}}
\newcommand{\LDf}{\LD^\text{(f)}}
\newcommand{\LDb}{\LD^\text{(b)}}
\newcommand{\LDfb}{\LD^\text{(fb)}}
\newcommand{\LDfbw}{\LD^\text{(fbw)}}
\newcommand{\Ws}{\mathcal{W}_\text{s}}
\newcommand{\Wu}{\mathcal{W}_\text{u}}
\newcommand{\Wsu}{\mathcal{W}_\text{s,u}}
\newcommand{\TSt}{\mathcal{T}}
\newcommand{\PO}{\mathcal{P}}
\newcommand{\weightingf}{\chi^\text{(f)}}
\newcommand{\weightingb}{\chi^\text{(b)}}
\newcommand{\weightingfb}{\chi^\text{(f,b)}}
\newcommand{\vtherm}{v_\text{therm}}
\newcommand{\Esaddle}{E^\ddagger}
\newcommand{\comment}[1]{\textsf{\textcolor{orange}{[#1]}}}
\newcommand{\sno}[1]{_\mathrm{#1}}
\newcommand{\no}[1]{\mathrm{#1}}
\newcommand{\acnew}[1]{\acfi{#1}\acused{#1}}
\renewcommand{\mathbf}[1]{\boldsymbol{#1}}

\begin{abstract}
Transition state theory formally provides a simplifying approach for
determining chemical reaction rates and pathways.
Given an underlying potential energy surface for a reactive system,
one can determine the dividing surface in phase space which separates 
reactant and product regions, and thereby also these regions.
This is often a difficult task, and it is especially demanding for
high-dimensional time-dependent systems or
when a non-local dividing surface is required.
Recently, approaches relying on Lagrangian descriptors have
been successful at resolving the dividing surface in some of
these challenging cases,
but this method can also be computationally expensive due to the
necessity of integrating the corresponding phase space function.
In this paper, we present an alternative method by which time-dependent,
\EDITS{locally} recrossing-free 
dividing
surfaces can be constructed without the
calculation of any auxiliary phase space function, but only from simple
dynamical properties close to the energy barrier.
\end{abstract}

\keywords{}
\maketitle

\section{Introduction}

Predicting the rate of a chemical reaction is a central problem in the study of
reaction dynamics. 
Their dynamics are governed by classical equations of
motion (EOM) driven by potentials whose primary nontrivial features
can be approximated locally
via rank-1 saddle points.
Transition state theory (TST)~\cite{pitzer,pechukas1981,truh79,truh85,%
truhlar91,truh96,truh2000,%
Komatsuzaki2001,Waalkens2008,hern08d,Komatsuzaki2010,hern10a,Henkelman2016}
then provides a powerful framework for predicting both reaction rates and pathways.
The theory is based on the identification of 
the dividing surface (DS) between reactant and product regions
associated with a given rank-1 saddle
which {\it ipso facto} defines these regions.
In order to obtain exact ---that is, not approximate---
reaction rates, the DS must be free of recrossings.
Otherwise, TST overestimates the reaction rate due to
the incorrect attribution of non-reactive flux to the reaction rate.
The central idea for constructing non-recrossing DSs is to
resolve the good action-angle variables in phase space
or the associated normally hyperbolic invariant manifold (NHIM)
via its stable and unstable
manifolds~\cite{Lichtenberg82,hern93a,Ott2002a,wiggins2013normally}.

Local approximations to the saddle geometry can be applied in the case of
low temperatures because passage over the rank-1 saddle is dominated
by trajectories that cross near it.
Harmonic approximations significantly reduce the computational effort because
the reaction rate can then be directly formulated from the properties of the
potential energy surface at the saddle.
For higher-energy contributions which can feel the anharmonic
potential in the vicinity of the saddle, 
normal form expansions~\cite{pollak78,pech79a,hern93b,hern94,Uzer02,Teramoto11,Li06prl,%
Waalkens04b,Waalkens13} can be used to construct non-recrossing
DSs to the desired accuracy.

A significant challenge arises if the DS is required for a 
potential energy surface whose reaction dynamics is determined by
a non-local region,
or that is time-dependent such as when external forces are present.
In recent years, this challenge has been successfully approached using
Lagrangian descriptors (LDs)~\cite{Mancho2010,Mancho2013,hern15e,hern16d},
which define a phase space function allowing for the identification of
both stable and unstable manifolds.
Despite their success in the application to non-autonomous
systems~\cite{hern16h,hern16i,hern17h}, 
we recognize that LD-based
methods can require a considerable amount of computational time
making it difficult to systems with large parameter sets or
many degrees of freedom.

In this paper, we introduce a method that efficiently computes \EDITS{discrete points on} the NHIM
from very simple dynamical properties close to the energy barrier.
A benefit of this algorithm lies in its avoidance 
of the calculation of any phase \EDITS{space function}
at the cost of the 
integration of a relatively small number of trajectories.
It converges exponentially
allowing for the determination of high-accuracy DSs with relatively small
computational effort.
\EDITS{The computational efficiency
in the determination of these high-accuracy points can be
extended through machine-learning algorithms
---e.g., based on neural networks \cite{hern18c}.
Specifically, a lower-resolution grid of points can be used
as a training set for a neural net thereby providing a
  non-iterative and smooth interpolation of the NHIM.
The recrossing-free dividing surface attached to this NHIM can then be
used in TST to compute reaction rates by propagating trajectories from
a suitably chosen initial distribution \cite{hern17h}.
Note that application of TST requires the knowledge of which time each
trajectory changes from reactant to product, and this is achieved with
the help of a recrossing-free time-dependent dividing surface
\cite{hern17h,hern18c,hern16a}.
}

The paper is organized as follows:
In Sec.~\ref{sec:LDs} we give a brief review of LDs and discuss a
method based on nested iterations with LDs to obtain points on the
NHIM.
We then introduce in Sec.~\ref{sec:binary_contraction} an efficient
binary contraction method that does not rely on LDs, but which
accomplishes the same goal.
Important aspects of the contraction method are discussed in
Secs.~\ref{sec:init} and~\ref{sec:error} in more detail.
In Sec.~\ref{sec:performance} we compare the performance of the binary
contraction method with the LD approach in a practical application.
\EDITS{The accuracy of the NHIM is also verified by showing
that points initially arbitrarily close to the NHIM deviate
away from it in time without recrossing it.}
\NEDITS{The limits and implications of this method in addressing 
higher-dimensional rank-1 saddles, multiple-ranked saddles,
roaming reactions and quantum effects are discussed in 
Sec.~\ref{sec:discussion}.}
Conclusions are drawn in Sec.~\ref{sec:conclusion}.

\section{Methods}
\label{sec:theory}

We describe a driven chemical reaction as a dynamical system 
with $d$ degrees of freedom 
of which only one is unstable and the remainder are stable.
The unstable degree of freedom is used to classify the reaction coordinate,
while we refer to the stable degrees of freedom that are coupled to the
dynamics of the reaction coordinate as the bath coordinates.
We model such a system by a Hamiltonian with a time-dependent rank-1
saddle driven by the underlying time-dependent potential.

When $d=1$, the dynamics in the vicinity of a saddle point 
can be surmised by the hyperbolic fixed
points in a moving frame.
Stable and unstable manifolds 
---that is, continuous sets of points in phase space where trajectories
either exponentially approach or leave the hyperbolic point, respectively---
can be attached to the hyperbolic points.
A critical observation lies in the fact
that dynamical propagation contracts the stable manifold
towards, and expands the unstable manifold away from the hyperbolic
fixed points.

In higher dimensions ($d>1$), the notion of a hyperbolic fixed point is
generalized by the $(2d-2)$-dimensional NHIM mentioned above, and
points on the stable and unstable manifolds exponentially approach to
or depart from the NHIM.
Once the NHIM is known, a recrossing-free DS with increased dimension
$2d-1$, which separates the $(2d)$-dimensional phase space into the
product and reactant region can easily be attached to the NHIM
\cite{hern17h}.

\subsection{Nested iterations with LDs}
\label{sec:LDs}

\begin{figure}
  \includegraphics[width=\columnwidth]{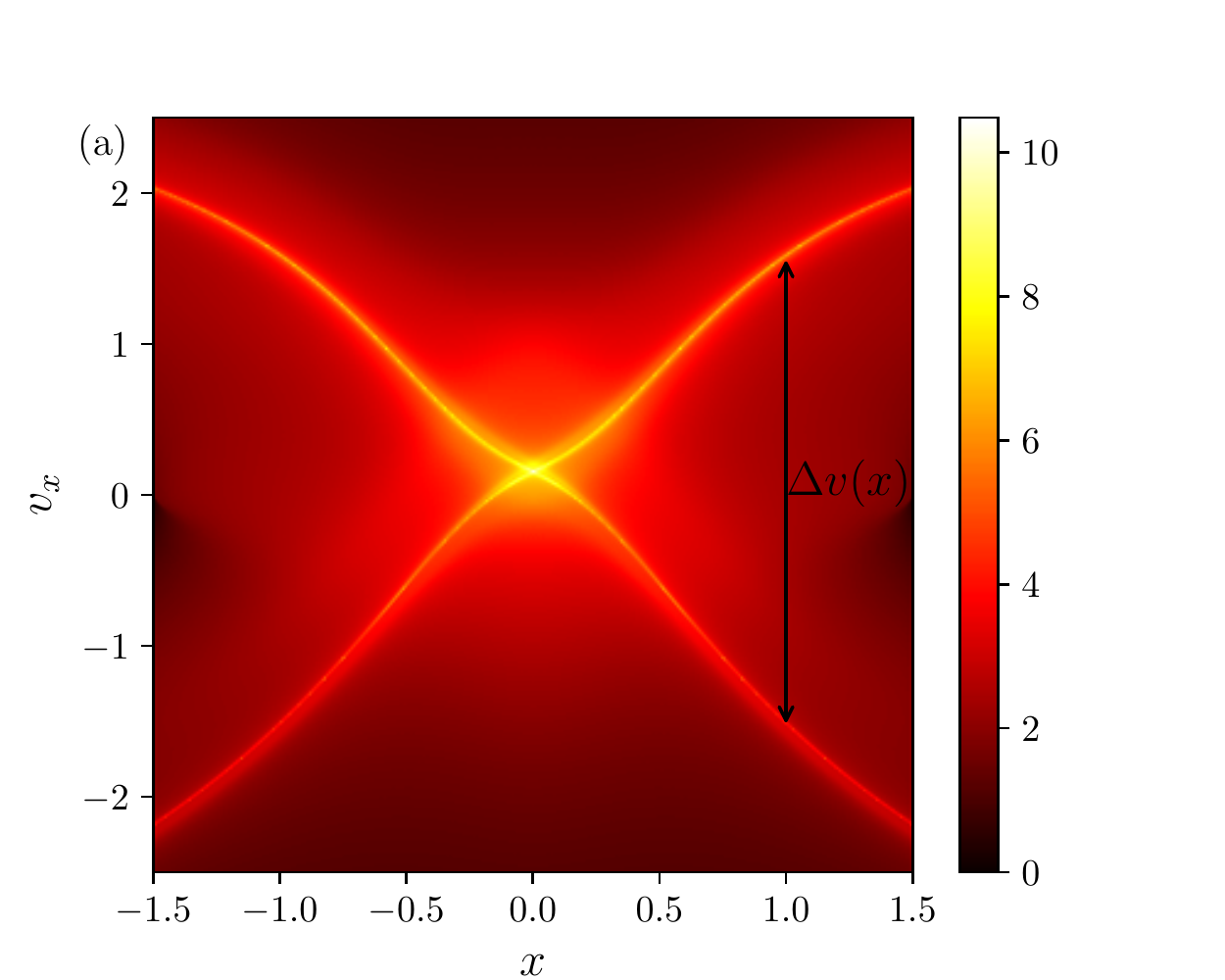}
  \includegraphics[width=0.9\columnwidth]{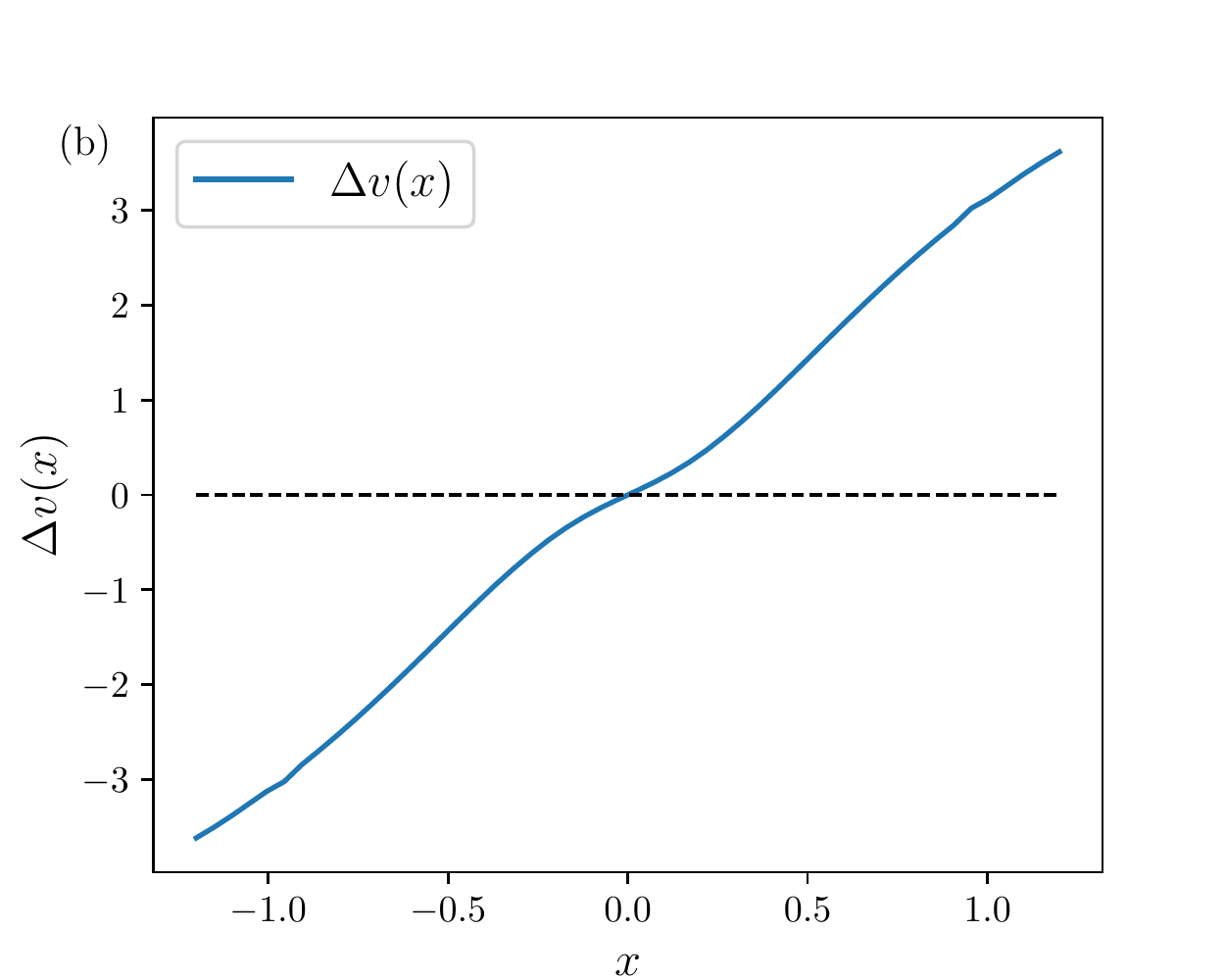}
  \caption{%
    (a)
    Two-dimensional LD plot in the $xv_x$-plane. The function $\Delta v(x)$ is
    given by the difference in $v_x$ between the extreme values for the forward
    and backward time component of the LD.
    (b)
    Plot of the function $\Delta v(x)$ for the figure in (a). The dotted
    line indicates the value zero of the function.
  }
  \label{fig:ld_delta}
\end{figure}

An LD is generally defined via the integral of a positive definite function 
along a trajectory within a given time interval $[t-\tau,~t+\tau]$.
Here, $t$ describes the time coordinate and $\tau$ is some positive value large
enough to cover the relevant dynamics.
The phase space function used to define the LD
is the velocity $\mathbf{v}$ of the particle, which makes the LD,
\begin{equation}
  \mathcal{L}(\mathbf{x}, \mathbf{v}, t) = \int_{t-\tau}^{t+\tau} ||
  \mathbf{v}(t^\prime) ||\, \mathrm{d}t^\prime\,,
\end{equation}
a measure of
the arc length of the corresponding 
trajectory~\cite{Mancho2013,hern15e,hern16a,hern16i}.
An example of a two-dimensional LD plot in the \mbox{$xv_x$-plane} computed
for the open, three-dimensional model system introduced later in
Sec.~\ref{sec:performance} is shown in Fig.~\ref{fig:ld_delta}(a).
The connection between the LD and the stable and unstable manifolds is
surprisingly simple~\cite{hern15e,hern16d}.
Since, the dynamics on these manifolds is extremal, said property is also true
for the LD.
To be precise, a particle on such a manifold approaches the fixed point
either in forward or in backward time.
Consequently, the LD reveals minimal values on each of the manifolds and
their intersection
---that is, the hyperbolic fixed point--- is a (local) minimum of the LD.

One can use this approach to construct an algorithm to find the intersection
of the closure of the stable and unstable manifold
---i.e.~the NHIM--- as follows.
First, at a given reaction coordinate $x$ and bath coordinates, 
one can perform an optimization in the reaction velocity $v_x$
extremizing the LD so as to obtain the associated positions of stable and
unstable manifolds [see Fig.~\ref{fig:ld_delta}(a)].
From this, one computes their difference $\Delta v(x)$ in $v_x$-space
as a function of the reaction coordinate $x$.
A one-dimensional root search for $\Delta v(x_0) = 0$
[see Fig.~\ref{fig:ld_delta}(b)]
yields the position $x_0$, where the stable and unstable manifold intersect,
i.e., a point on the NHIM.
This can be repeated for a set of bath coordinates and time to obtain
the time-dependent NHIM up to the desired resolution.

Although one can readily identify the intersection of the manifolds in
Fig.~\ref{fig:ld_delta}(a), the construction of a high
accuracy image requires the computation of tens of thousands of
trajectories and is only useful for demonstration purposes or
to determine a rough estimate for the location of the NHIM.
Meanwhile, the computation of a single value of $\Delta v(x)$
---i.e., a search of the manifolds for a fixed reaction coordinate---
requires the computation of several LDs
because extremal optimizations are in general
iterative processes.
This is especially taxing as one considers the fact that root finding methods,
in and of themselves, are also iterative and require multiple values of 
$\Delta v(x)$
to provide accurate results.
Doing so, provides a multiplicative factor to the required amount
of computations.
For multidimensional systems, these nested iterations have to be
performed several times over large sets of bath coordinates, 
and this may
easily result in the integration of several millions of trajectories.

\subsection{Binary contraction method}
\label{sec:binary_contraction}

\begin{figure}
 \includegraphics[width=0.9\columnwidth]{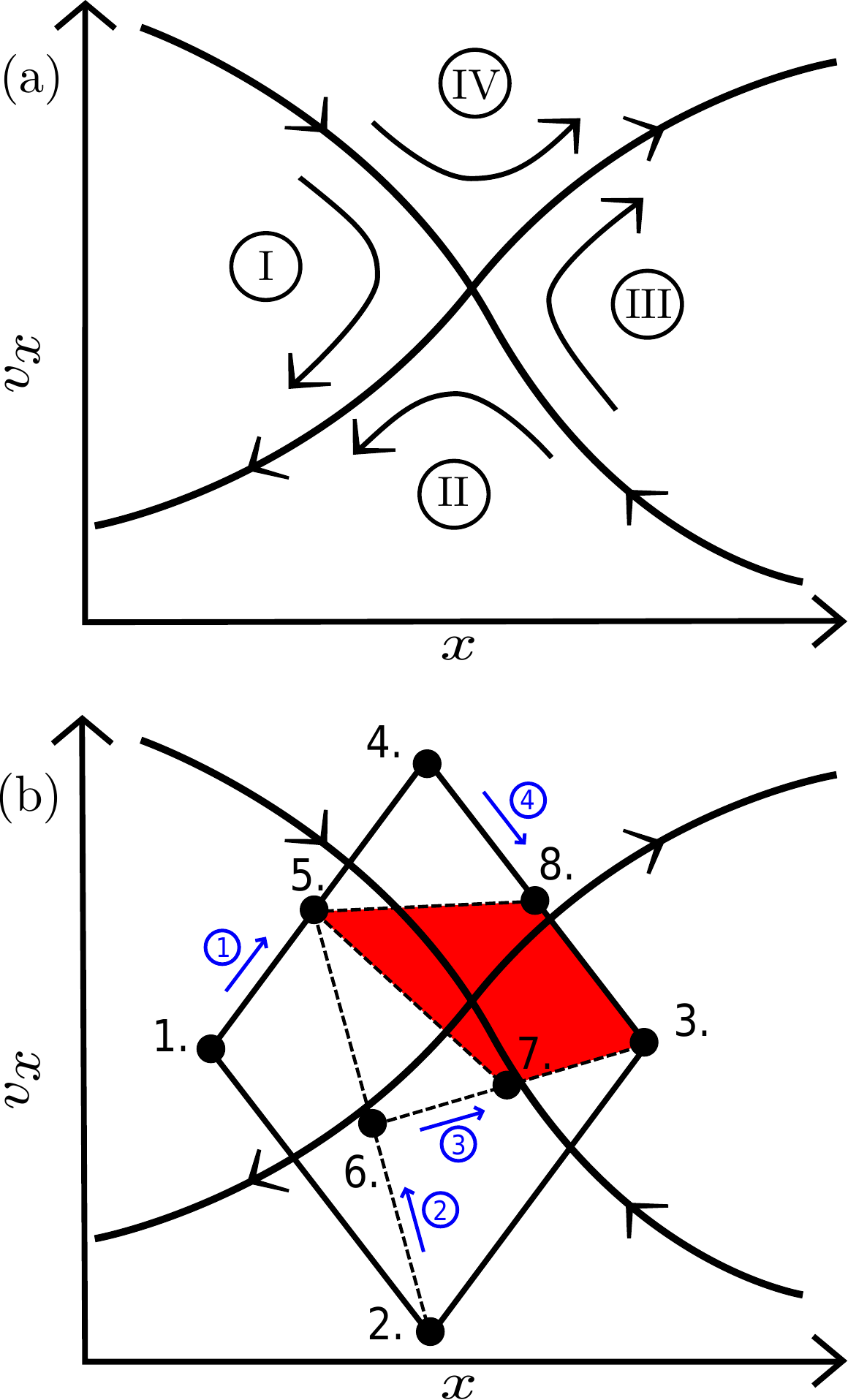}
\caption{%
(a)
View of the phase space in reaction coordinate and velocity. Stable
and unstable manifold intersect and divide the plane in four regions
marked by (I)--(IV).  The dynamics in a region is dictated by the
dynamics on the manifolds that form the region's boundary. The arrows
indicate the general path of trajectories in a positive time direction.
(b)
The intersection of the manifolds is found by contracting a quadrangle with a
binary search algorithm applied to all four edges consecutively. The highlighted
quadrangle is the result of contracting all edges of the quadrangle with the
vertices 1, 2, 3, and 4 once in a counter-clockwise manner.  The arrows
with encircled numbers indicate the order and direction how vertices
are replaced.  The procedure can be iterated to obtain the
intersection point with the desired accuracy.
}
\label{fig:manifolds}
\end{figure}

In this section, we provide an alternative method that does not explicitly
search for the stable and unstable manifolds, 
and thereby avoids the nested
iterations of the LD approach.
To accomplish this, we assume that the saddle region 
is an appropriate interval in
the reaction coordinate $[x_1, x_2]$ that covers the relevant dynamics of the
time-dependent saddle.
\EDITS{It can be found by
sampling LDs in the region of interest,
and has the structure shown in Fig.~\ref{fig:ld_delta}(a).}
In doing so, we are able to discriminate between four types of
trajectories tied to one of four regions in phase space
as shown in Fig.~\ref{fig:manifolds}(a).
\EDITS{Trajectories in regions (II) and (IV) are characterized as
  reactive, going from reactants to products and vice versa, respectively.
  Trajectories in regions (I) and (III) are non-reactive, 
  remaining as products to products and reactants to reactants, respectively.
  Thus, every trajectory can be assigned to one of the four regions by
  integrating it forward and backward in time.
  While trajectories near the stable manifolds can spend
  a long time in the vicinity of the NHIM, 
they will eventually leave the barrier region when they are
  integrated sufficiently long.
  For orbits extremely close to the stable manifolds, the classification
  to the regions (I)-(IV) may nevertheless depend on rounding errors or numerical
  errors of the integrator, which means that the stable and unstable
  manifolds and thus the crossing point on the NHIM can only be determined
  with finite numerical accuracy. 
For the benchmark example of this work, the accuracy of the integrator
was sufficient for the determination of the four regions as illustrated
in Sec.~\ref{sec:tracing}.}

Our binary contraction method is based on the observation that in the
immediate vicinity of the hyperbolic point, 
the reactive and non-reactive regions
are arbitrarily close to each other but still separated by
stable and unstable manifolds.
This is illustrated in \FIG~\ref{fig:manifolds}(a). 
The manifolds
associated with the NHIM determine the particle dynamics in the saddle
region due to the asymptotic behavior of the related trajectories.
More precisely, these regions determine 
which directions a particle enters and leaves the saddle region.

Similar to the LD method, we integrate the trajectories forward and
backward in time. \EDITS{However, here the propagation of the trajectories does
not occur for a fixed amount of time, rather
each of the integrations in
forward and backward time
is stopped when the particle leaves the saddle region.}
This allows us to determine which of the four regions corresponds
to the given initial phase space point.
Using this information, we can construct a quadrangle in the
\mbox{$xv_x$-plane}, with each vertex in one of the four regions.
The algorithm is then divided into five steps [see \FIG~\ref{fig:manifolds}(b)]:
\begin{enumerate}
\item
Construct a quadrangle with each of its vertices in one of the four regions
(I)--(IV).
\item
Determine the midpoint between two adjacent vertices of the quadrangle.
\item
Determine {which region}
the midpoint corresponds to by
integrating the dynamics in forward and backward time.
\item
Use the new vertex to replace the vertex in the same region.
(If the region identified for the midpoint is not identical 
to either of the original
points, perform an error correction as described in \SEC~\ref{sec:error}.)
\item
Repeat steps 2--4 for all edges, e.g., in a counter-clockwise manner
as in Fig.~\ref{fig:manifolds}, until the longest edge of the
quadrangle is below a desired error tolerance.
\end{enumerate}

As implemented in our scheme, we walk through all the edges counterclockwise
and successively in step~5.
Alternative walks through the edges could provide faster or slower
convergence, and the optimization of this scheme would be of possible 
interest to future work.

In \FIG~\ref{fig:manifolds}(b), we illustrate these steps for the initial
quadrangle with vertices $\lbrace 1,2,3,4 \rbrace$.
First, our algorithm contracts vertices 1 and 4, leading to the new quadrangle
$\lbrace 5,2,3,4 \rbrace$.
It is then applied to vertices 5 and 2 to create $\lbrace 5,6,3,4\rbrace $.
In the next step, the new vertex 6 (and not one of the old vertex points
as in the two previous steps) is replaced with vertex 7 resulting in
$\lbrace 5,7,3,4\rbrace $.
After step~4, the highlighted quadrangle $\lbrace 5,7,3,8 \rbrace $ is created.
Since this process is similar to a binary search algorithm, we call this
method binary contraction.

Repeating this for all lines on the quadrangle ensures the
reduction of mutual phase space distances between the vertices.
Since reducing these distances \EDITS{in the $xv_x$-plane} to an arbitrarily small size is only
possible in the immediate vicinity of the hyperbolic point, we know that the quadrangle
converges towards it. \EDITS{By computing the geometric center of the quadrangle in that plane}, we have found the position of the hyperbolic
point with an accuracy proportional to the size of the final quadrangle.
\EDITS{By applying this method on a $xv_x$-plane of the reaction coordinates 
in a higher dimensional system, 
one can construct the NHIM $(x^{\mathrm{NHIM}}(\mathbf{y},
\mathbf{v}_y, t_0), v_x^\mathrm{NHIM}(\mathbf{y}, \mathbf{v_y}, t_0))$
for any suitable set of bath coordinates $(\mathbf{y},\mathbf{v_y})$
and initial time $t_0$.  As the contraction still takes place on a
two-dimensional plane, the efficiency of the method is not impacted by
additional dimensions in phase space.}

An efficient initialization of the quadrangle in step 1, as well as
the error correction in step 4 will now be discussed in detail in
Secs.~\ref{sec:init} and~\ref{sec:error}, respectively.

\subsubsection{Exception handling in initialization}
\label{sec:init}

An important prerequisite of the binary contraction method is the identification of
good initial conditions for the iteration.
To construct a valid quadrangle, we rely on the observation that
the shape of the cross which divides the regions in the $xv_x$-plane
is robust in the vicinity of the hyperbolic fixed point.
\EDITS{The construction of the NHIM can be visualized readily 
with the help of an LD 
plot such as that shown in Fig.~\ref{fig:ld_delta}(a).}
One \EDITS{first} constructs a quadrangle by guessing
a phase space coordinate $(x_0, v_{x_0})$ in the vicinity of the NHIM.
Then one constructs the vertices with constant shifts in the $xv_x$-plane
$\Delta x$ and $\Delta v_x$, such that $(x_0 \pm \Delta x , v_{x_0})$ and
$(x_0, v_{x_0} \pm \Delta v_x)$ form the vertices of the quadrangle.

For the initial set of selected vertices, 
we integrate the trajectories and compare their
types to see if the construction of the required quadrangle was successful
in the sense that there is a vertex in each of the 4 regions.
If this is true, the iteration will start at this point.
However some constructions may fail because they rely on both the
accuracy of the guess and the size of the shifts.
To mitigate this issue, one may increase shift sizes and try again.
In addition, one can implement a bookkeeping method to combine valid
vertices constructed from different shift sizes.
Optimally, using a light-weight extra\-polation method to determine an
accurate guess from previous results would allow for a quick construction.
This is especially efficient if we construct the NHIM in higher
dimensions from a few already known points.
In practice, a combination of both extrapolation and managed
variation of shift sizes have proven to be highly effective for 
constructing the initial quadrangle.

One could also try to use an LD-based approach to determine points on
the stable and unstable manifolds from which one can interpolate a
valid quadrangle.
However, because of the efficiency of the iteration
in the binary contraction method, 
it would be ill-advised to use an initialization method that
is computationally taxing in comparison, as the latter could dominate the
runtime and overall cost of the computation.

\subsubsection{Increasing convergence for edge cases}
\label{sec:error}
\begin{figure}
   \includegraphics[width=0.9\columnwidth]{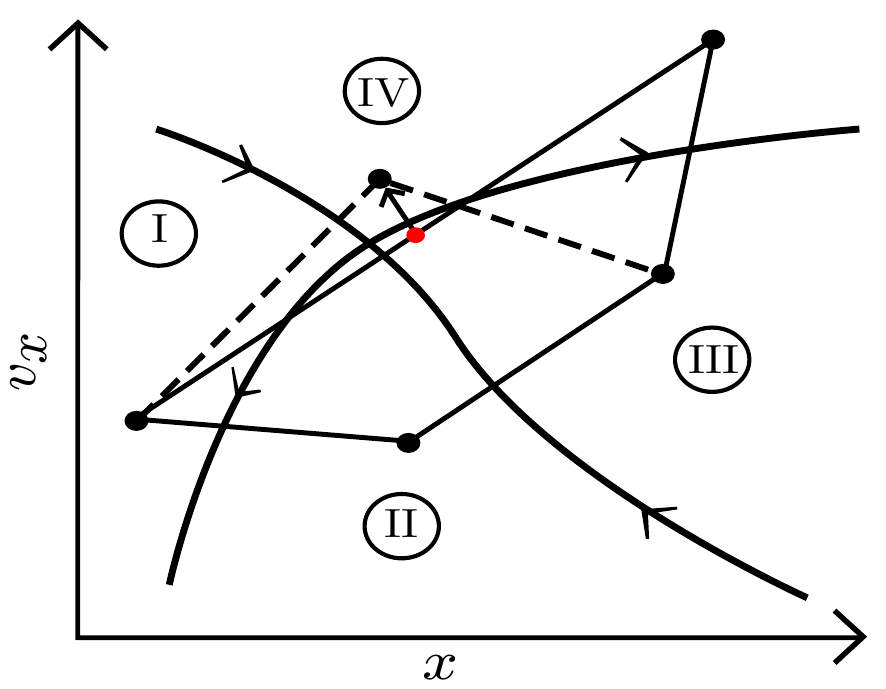}%
  \caption{%
  Edge case where a convergence failure may occur. The connecting line
  between vertices in regions (I) and (IV) crosses into other regions.
  The highlighted midpoint is detected to be in region (III) instead of
  (I) or (IV), and a new vertex is searched in the direction orthogonal
  to the connecting line. The dotted lines highlight the new quadrangle
  after a successful correction.}\label{fig:edgecase}
\end{figure}

Depending on the choice of the underlying EOM and plane, one may run
into situations where the borders of the regions form a heavily
distorted cross.
In situations like these, it is possible that the quadrangle contracts
into an area where the intersection is not present, see Fig.~\ref{fig:edgecase}.
For an iteration that is completely inclusive
---that is, one that contracts the quadrangle into a subset of itself---
this may lead to convergence problems.
This issue arises when the midpoint of a line is not contained in
either of the regions of the original vertices, as this is the only
situation where the quadrangle does not contract further.
In that case, we allow the quadrangle to expand outward in the
direction where the failure was discovered.
The expansion is controlled such that the possible contraction of the
iteration dominates.

Using this procedure, the quadrangle not only contracts
around the intersection of the region, but also tumbles along the
stable and unstable manifolds to actively center itself around the
intersection.
Although this procedure adds computational complexity to the
iteration, we have observed that, on average, there is no noticeable
increase in the computational time required.
On the contrary, it even accelerates the computation for situations where
convergence might not have been reached otherwise
because it opens the quadrangle to more equally wrap around the 
targeted fixed point.

\section{Results}
\label{sec:performance}
\subsection{\EDITS{Tracing the motion of the NHIM for fixed bath coordinates}}
\label{sec:tracing}
In this section, we will compare the performance of the LD-based method with
nested iterations to the binary contraction method by \EDITS{computing several
points on the NHIM and comparing the average number of necessary trajectories to
compute a single point as a function of the algorithm's tolerance. This is
accomplished by} applying both \EDITS{methods} to a three-dimensional model
system with a time-dependently moving rank-1 saddle. To this end, we use a
three-dimen\-sional extension of the two-dimensional
time-dependent potential introduced in \REF~\cite{hern17h}
\begin{align}
    V(x,y,z,t) &= E_b \exp{\left(-a {\left[ x - \hat x \sin{(\omega_x t)}
    \right]}^2\right)} \nonumber \\
    &+ \frac{\omega_y^2}{2} {\left(y - \frac{2}{\pi}
    \arctan{(2x)}\right)}^2 \nonumber \\
    &+ \frac{\omega_z^2}{2} {\left(z - \frac{2}{\pi}
    \arctan{(2x)}\right)}^2\ ,
\label{eq:pot}
\end{align}
and we use parameters in simulation units $E_b = 2$, $a = 1$,
$\hat x = 0.4$, $\omega_x = \pi$, $\omega_y = 2$, and $\omega_z = 1$.
With these parameters, the potential has a periodicity of $T = 2$.
The algorithms are set to find a point on the NHIM for fixed bath coordinates 
$y = z = v_y = 0$ and $v_z = 0.5$, for 200 \EDITS{equidistant, initial} time
coordinates 
\EDITS{during one oscillation period, thereby tracing the NHIM's
time-dependent motion.}
\EDITS{Note that this tracing
of the NHIM is not, in general, a trajectory of the system, since the results of
the algorithms are not propagated by the dynamics of the system to find further
points on the NHIM. 
Each point is computed separately. 
This allows us to
control the accuracy of the data, as well as the specific bath coordinate for
which the point is computed.}
\begin{figure}[t]
 \includegraphics[width=0.9\columnwidth]{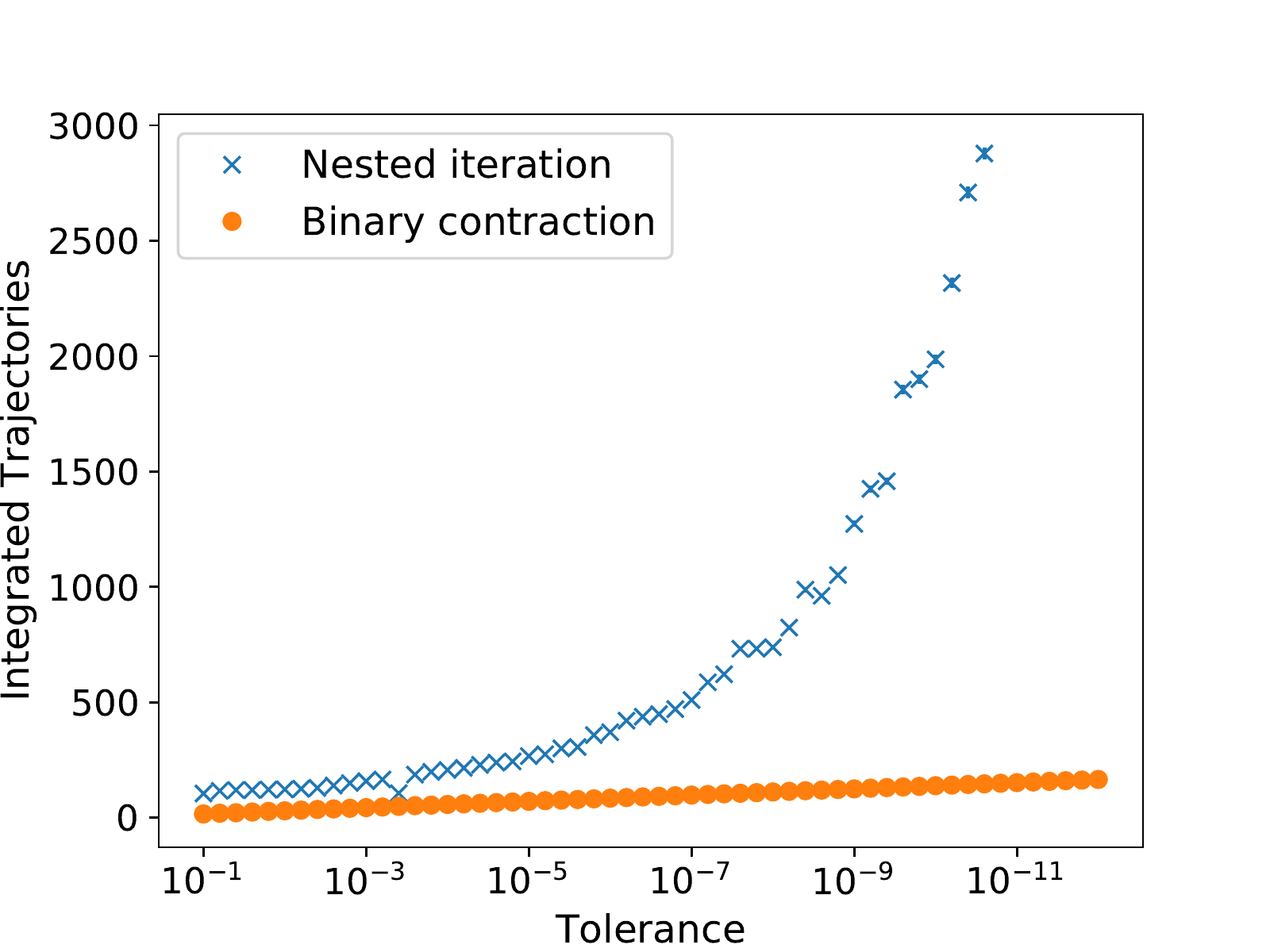}
\caption{Number of integrated trajectories per hyperbolic fixed point
  found as function of the error tolerance of the algorithms. The
  difference in the scaling behavior is evident.}\label{fig:performance}
\end{figure}

The results in \FIG~\ref{fig:performance} show that the binary
contraction method requires far 
\EDITS{fewer} trajectories to reach a given
tolerance.
Especially for very low tolerances, i.e., for high accuracy, it is
highly superior in terms of performance.
The difference in scaling is also quite evident, as the contraction
exhibits an exponential convergence to the hyperbolic fixed point,
while the performance of the nested iterations is clearly worse, even
for large values of the tolerance.
The fast convergence of the binary contraction method is evident from
the fact that it divides the lengths of the quadrangle's sides by
roughly two for every successful iteration, therefore contracting the
quadrangle exponentially fast.
\begin{figure}[t]
 \includegraphics[width=\columnwidth]{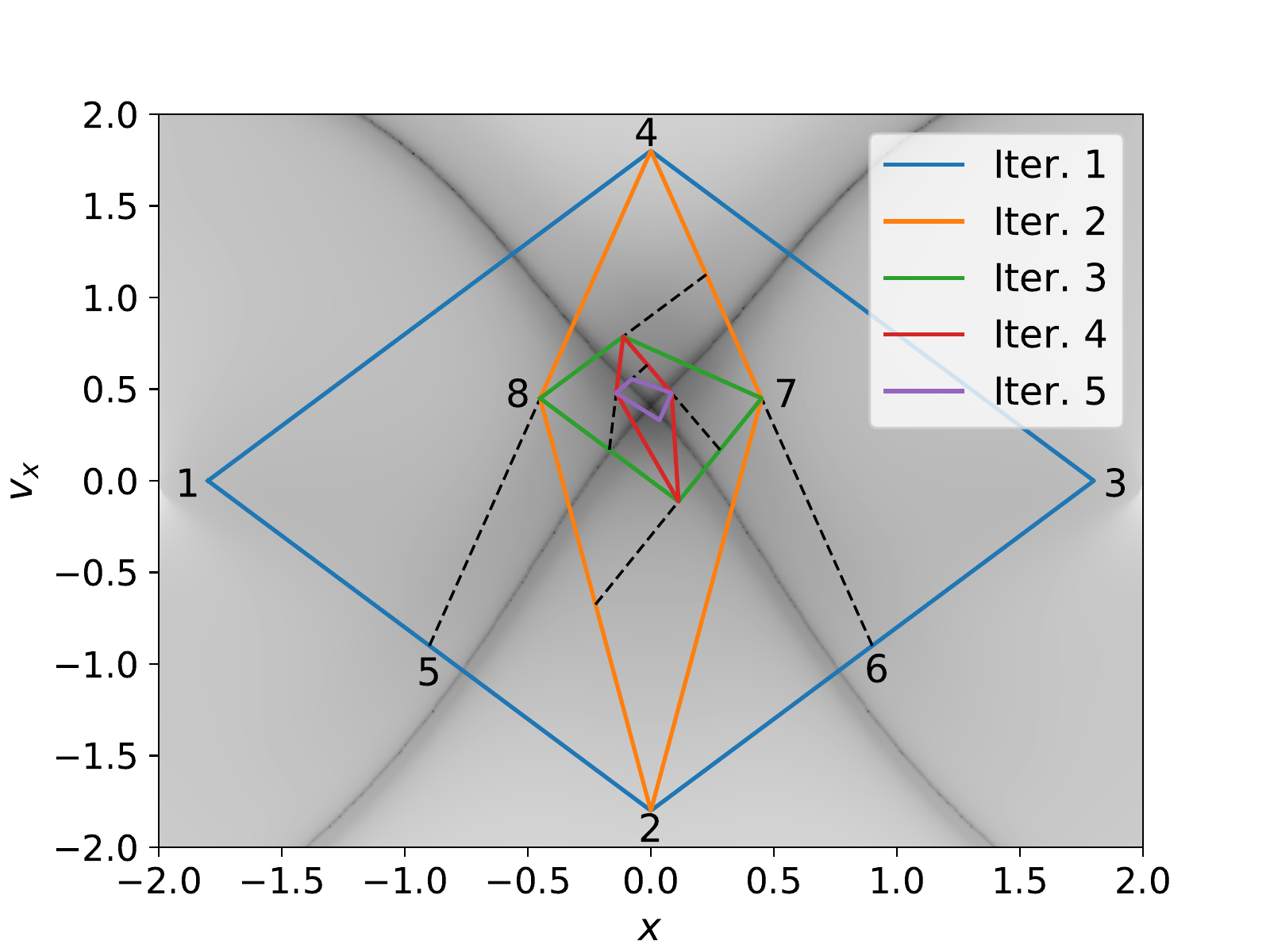}
\caption{%
The binary contraction in action for the model system with the
time-dependent potential given in Eq.~\eqref{eq:pot}. One iteration
corresponds to one counter-clockwise revolution around the quadrangle,
contracting each side once, starting from the vertex line $\lbrace 1,2\rbrace $
of the quadrangle $\lbrace 1,2,3,4 \rbrace $. In the next iteration the 
quadrangle $\lbrace 2,7,4,8 \rbrace $ is obtained. The LD values are
shown through shading in the background, where one can clearly discern
a cross which corresponds to the stable and unstable manifolds in the 
darkest shade. The
quadrangles rapidly contract towards their intersection.
}
\label{fig:rauten}
\end{figure}
This is further illustrated in Fig.~\ref{fig:rauten}, where we apply
the method to the model system with the time-dependent
potential~\eqref{eq:pot} for a specific choice of the bath coordinates
and time.
[Note that in this figure we use a different set of bath coordinates
and time compared to Fig.~\ref{fig:ld_delta}(a).]
The quadrangles of the first and second iteration are marked by
numbers 1--8, and the dashed lines complete intermediate quadrangles.
The quadrangles converge rapidly towards the intersection of the
stable and unstable manifolds as marked by the largest values of the 
LD shown in the darkest shade of gray.

\subsection{\EDITS{Propagation of trajectories initially in 
close vicinity to the NHIM}}
\EDITS
{
Using our method to efficiently compute high-accuracy points on the NHIM for any
suitable bath coordinate and time, we investigate how trajectories 
deviate from the NHIM when they are started in its close vicinity. 
In this section, we present the results of an investigation for
the model system defined in Eq.~\eqref{eq:pot} as it is
sufficient to illustrate the behavior even for systems with
higher dimensionality as long as our earlier stated assumptions
are satisfied. 
}
\begin{figure}
    \includegraphics[width=\columnwidth]{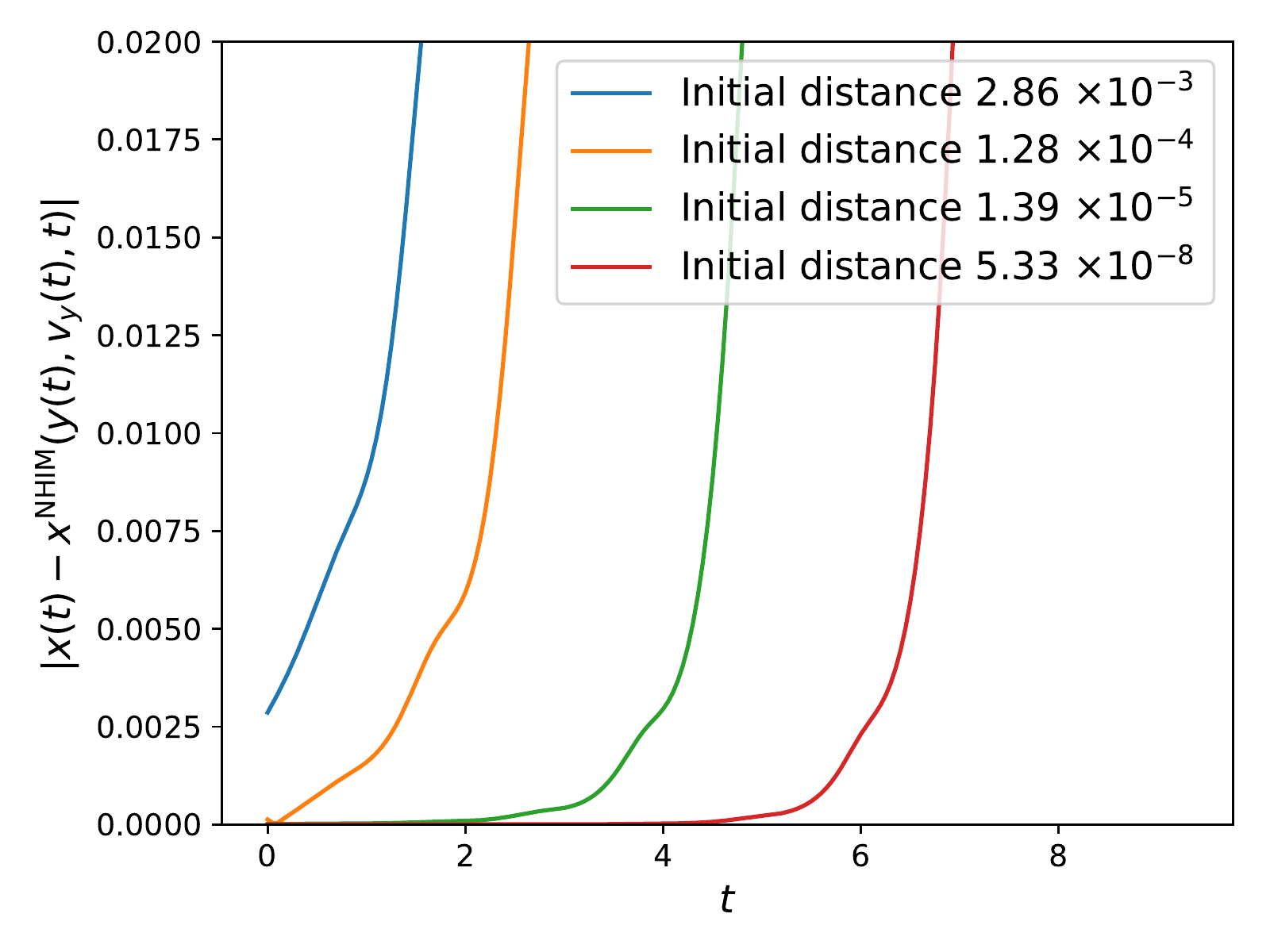}
    \caption { \EDITS {
        The $x$-coordinate distance of trajectories from the NHIM over
        time. The trajectories were propagated from the close vicinity of the
        NHIM
with the initial position at increasing distances from the NHIM
as listed in the legend and seen in the curves from left to right.
        Trajectories starting closer to the NHIM stay in its vicinity
        for a longer amount of time before deviating from it. 
    } }
    \label{fig:nhimtraj}
\end{figure}
\EDITS{The binary contraction was performed to find the
reaction coordinate $(x^\mathrm{NHIM}, v_x^\mathrm{NHIM})$ of the NHIM for 
a given
bath coordinate. 
The results of the contraction, performed for
several different accuracies, are used as initial conditions for trajectories
starting very close to the NHIM.}

\EDITS{The properties of the resulting trajectories are shown
in Fig.~\ref{fig:nhimtraj}. 
Trajectories initially in close vicinity to the NHIM deviate 
exponentially from it, as expected.
Those trajectories that started closer to the NHIM stay in
its vicinity for longer time. 
When trajectories were initiated on the numerically computed
NHIM, the situation was rather different in that they would
also eventually deviate from the NHIM.
In order to limit this deviation, the accuracy of the calculation had
to be increased by continuing the iterative process of the contraction
until the longest distance between adjacent vertices of the quadrangle
was below $10^{-12}$.
It is important to note that this accuracy only applies to 
the discrete dynamics of the numerically integrated system. 
The analytical accuracy of the trajectories
is bound by the accuracy of the integration method used, as well as the size
of the discrete time step. 
Nevertheless, for the discrete dynamics, 
this degree of accuracy can be achieved as long as
sufficient numerical stability is provided.
Consequently, the binary contraction method can indeed be used to
obtain stable and accurate estimates of the NHIM
up to the desired numerical accuracy.
}

\section{Discussion}\label{sec:discussion}

This work has been restricted to rank-1 saddles
primarily because the construction of the time-dependent
transition state trajectory remains a challenge even for
this case.
Much of the early work \cite{dawn05a,dawn05b,hern06d}
on the transition state trajectory
utilizing perturbation theory presumes
such rank-1 saddles but is restricted by its radius of
convergence.
The more recent use of the LD was meant to get around this
problem but it can be challenging when there is higher 
dimensionality
\cite{Mancho2010,Mancho2013,hern15e,hern16d,hern16h,hern16i,hern17h}.
Below, we discuss the extent to which the binary contraction method
addresses this challenge, and the implications on activated
reactions with more general features.

\NEDITS{
The simple geometric structure drawn in, for example, Fig.~\ref{fig:manifolds}
may appear to be a challenge to the binary contraction method for
rank-1 barriers with multidimensional orthogonal coordinates.
However, regardless of dimensionality
the method contracts onto a quadrangle whose vertices reside on a
two-dimensional plane via the midpoints between the vertices.
The number of
trajectories necessary to contract the quadrangle to the desired size does not
increase with the dimensionality of the problem. 
However, the complexity of the
equations of motion that may increase in higher dimensional systems will
increase the computational cost of computing these trajectories.
Additionally, one needs to keep in mind that even though the complexity of the
contraction for a single point on the NHIM scales modestly with the dimensionality of
the equations of motion, the number of points needed to characterize the NHIM
for a given system may scale exponentially with its dimensionality, depending on
how one chooses to construct the DS.
}

\NEDITS{
Although we have been successful in applying the binary contraction method 
to a system with a
rank-1 saddle potential, we have yet to see how well it can be applied to
systems with higher ranked saddle potentials, such as the systems investigated in
Refs.~\citenum{jaffe10, jaffe11, komatsuzaki13a}, where the classification of
trajectories is not as straightforward as for rank-1 saddle potentials
because the characterization of the dynamics does not necessarily 
contract to the requisite quadrangle.
Thus it remains to be seen if this approach can be extended to higher-rank
saddles.
}

\NEDITS{
Another challenge to this and related approaches lies in its possible application
to roaming pathways for which no clear saddle 
exists \cite{bowman04a, hern13e, wiggins14a, bowman2011c,bowman14a}.
However, if a NHIM exists for such a reaction, then that NHIM
would by definition not be associated to a rank-1 saddle potential. 
This makes the identification of trajectories for the 
binary contraction even more of a challenge because the requisite
quadrangle does not exist.
Thus the future use of a binary contraction method to address roaming
reactions would require 
a classification scheme to identify which trajectories 
---on a contracted two-dimensional surface--- are reactive
or not.
}

\NEDITS{
Finally, an additional concern for this and other related approaches addressing
chemical reactions using transition state theory is whether quantum effects
provide a significant correction \cite{mill74,truh82}.
Indeed, one of the constructions of semiclassical transition state theory
has relied on the construction of the good-action variables that
underlie the NIHM \cite{mill77,mill90b,hern93b,hern94,Stanton2011,Stanton2012}.
Thus, one possible new direction beyond this work would involve the use of the
binary contraction method to construct not use the NIHM but also an underlying
coordinate space that represents the nonseparable but integrable coordinates
needed to construct a semiclassical rate formula.
However, this would only be necessary if the chosen system exhibits significant
quantum effects. 
The present work is restricted to fully classical treatments of driven chemical
reactions.
Such dynamics arises frequently, particularly when heavy molecular motions 
occur in solution.
}

\section{Conclusion}
\label{sec:conclusion}

We have introduced an exponentially converging algorithm that finds
hyperbolic fixed points, i.e.\ the NHIM for non-autonomous systems
with rank-1 saddles.
The binary contraction method achieves the high efficiency by avoiding
the explicit search for stable and unstable manifolds in nested
iterations with LDs.
The method relies on the identification of the saddle region, as well
as the ability to describe the system in coordinates where the stable and
unstable manifold intersect within two-dimensional planes in phase space.

\EDITS{By independently performing the algorithm on several two-dimensional
planes one can map the NHIM via the bath coordinates of the system.} 
We have discussed ideas to make this method reliable for difficult
phase space structures, and methods for efficient initialization even
in higher dimensions.

\EDITS{We have confirmed the relative stability
of the NHIM computed numerically from the binary contraction method.
Specifically, trajectories initiated on the 
time-dependent NHIM remain in its
vicinity over time subject to the accuracy of the 
initial computation of the NHIM.
}
The binary contraction method presented here is 
a useful advance to the existing set of methods for 
the comprehensive handling of driven chemical reactions.
Using TST, \NEDITS{fluxes and}
rates in such systems can be obtained by propagating a
large ensemble of trajectories and determining the time when each trajectory
crosses a time-dependent DS~\cite{hern17h}.
The binary contraction method provides, in a first step, a large set
of points on the NHIM, which can then be used, in a second step, to
construct a time-dependent and \EDITS{locally} recrossing-free DS by interpolating the
discrete points on the NHIM.
\NEDITS{As with the other approaches ---listed in the discussion
in Sec.~\ref{sec:discussion},--- 
it remains challenging to address high dimensionality in the
degrees of freedom orthogonal to the reaction coordinate.}
This
\NEDITS{could perhaps be resolved}
using machine learning methods, such as neural
networks, as demonstrated in Ref.~\citenum{hern18c}.

\section*{Acknowledgments}
The German portion of this collaborative work was partially supported
by Deutsche Forschungsgemeinschaft (DFG) through Grant No.\ MA1639/14-1.
The US portion was partially supported by the National Science Foundation (NSF)
through Grant No.~CHE 1700749.
AJ acknowledges the Alexander von Humboldt Foundation, Germany, for
support through a Feodor Lynen Fellowship.
MF is grateful for support from the Landesgraduiertenf\"orderung of
the Land Baden-W\"urttemberg.
This collaboration has also benefited from support by the European
Union's Horizon 2020 Research and Innovation Program under the Marie
Sklodowska-Curie Grant Agreement No.~734557.

\bibliography{paperq08}

\begin{thebibliography}{52}%
\makeatletter
\providecommand \@ifxundefined [1]{%
 \@ifx{#1\undefined}
}%
\providecommand \@ifnum [1]{%
 \ifnum #1\expandafter \@firstoftwo
 \else \expandafter \@secondoftwo
 \fi
}%
\providecommand \@ifx [1]{%
 \ifx #1\expandafter \@firstoftwo
 \else \expandafter \@secondoftwo
 \fi
}%
\providecommand \natexlab [1]{#1}%
\providecommand \enquote  [1]{``#1''}%
\providecommand \bibnamefont  [1]{#1}%
\providecommand \bibfnamefont [1]{#1}%
\providecommand \citenamefont [1]{#1}%
\providecommand \href@noop [0]{\@secondoftwo}%
\providecommand \href [0]{\begingroup \@sanitize@url \@href}%
\providecommand \@href[1]{\@@startlink{#1}\@@href}%
\providecommand \@@href[1]{\endgroup#1\@@endlink}%
\providecommand \@sanitize@url [0]{\catcode `\\12\catcode `\$12\catcode
  `\&12\catcode `\#12\catcode `\^12\catcode `\_12\catcode `\%12\relax}%
\providecommand \@@startlink[1]{}%
\providecommand \@@endlink[0]{}%
\providecommand \url  [0]{\begingroup\@sanitize@url \@url }%
\providecommand \@url [1]{\endgroup\@href {#1}{\urlprefix }}%
\providecommand \urlprefix  [0]{URL }%
\providecommand \Eprint [0]{\href }%
\providecommand \doibase [0]{http://dx.doi.org/}%
\providecommand \selectlanguage [0]{\@gobble}%
\providecommand \bibinfo  [0]{\@secondoftwo}%
\providecommand \bibfield  [0]{\@secondoftwo}%
\providecommand \translation [1]{[#1]}%
\providecommand \BibitemOpen [0]{}%
\providecommand \bibitemStop [0]{}%
\providecommand \bibitemNoStop [0]{.\EOS\space}%
\providecommand \EOS [0]{\spacefactor3000\relax}%
\providecommand \BibitemShut  [1]{\csname bibitem#1\endcsname}%
\let\auto@bib@innerbib\@empty
\bibitem [{\citenamefont {Pitzer}\ \emph {et~al.}(1962)\citenamefont {Pitzer},
  \citenamefont {Smith},\ and\ \citenamefont {Eyring}}]{pitzer}%
  \BibitemOpen
  \bibfield  {author} {\bibinfo {author} {\bibfnamefont {K.~S.}\ \bibnamefont
  {Pitzer}}, \bibinfo {author} {\bibfnamefont {F.~T.}\ \bibnamefont {Smith}}, \
  and\ \bibinfo {author} {\bibfnamefont {H.}~\bibnamefont {Eyring}},\
  }\href@noop {} {\emph {\bibinfo {title} {The Transition State}}},\ Special
  Publ.\ (\bibinfo  {publisher} {Chemical Society},\ \bibinfo {address}
  {London},\ \bibinfo {year} {1962})\ p.~\bibinfo {pages} {53}\BibitemShut
  {NoStop}%
\bibitem [{\citenamefont {Pechukas}(1981)}]{pechukas1981}%
  \BibitemOpen
  \bibfield  {author} {\bibinfo {author} {\bibfnamefont {P.}~\bibnamefont
  {Pechukas}},\ }\href@noop {} {\bibfield  {journal} {\bibinfo  {journal}
  {Annu. Rev. Phys. Chem.}\ }\textbf {\bibinfo {volume} {32}},\ \bibinfo
  {pages} {159} (\bibinfo {year} {1981})}\BibitemShut {NoStop}%
\bibitem [{\citenamefont {Garrett}\ and\ \citenamefont
  {Truhlar}(1979)}]{truh79}%
  \BibitemOpen
  \bibfield  {author} {\bibinfo {author} {\bibfnamefont {B.~C.}\ \bibnamefont
  {Garrett}}\ and\ \bibinfo {author} {\bibfnamefont {D.~G.}\ \bibnamefont
  {Truhlar}},\ }\href@noop {} {\bibfield  {journal} {\bibinfo  {journal} {J.
  Phys. Chem.}\ }\textbf {\bibinfo {volume} {83}},\ \bibinfo {pages} {1052}
  (\bibinfo {year} {1979})}\BibitemShut {NoStop}%
\bibitem [{\citenamefont {Truhlar}\ \emph {et~al.}(1985)\citenamefont
  {Truhlar}, \citenamefont {Issacson},\ and\ \citenamefont {Garrett}}]{truh85}%
  \BibitemOpen
  \bibfield  {author} {\bibinfo {author} {\bibfnamefont {D.~G.}\ \bibnamefont
  {Truhlar}}, \bibinfo {author} {\bibfnamefont {A.~D.}\ \bibnamefont
  {Issacson}}, \ and\ \bibinfo {author} {\bibfnamefont {B.~C.}\ \bibnamefont
  {Garrett}},\ }\enquote {\bibinfo {title} {Theory of chemical reaction
  dynamics},}\ \ (\bibinfo  {publisher} {CRC Press},\ \bibinfo {address} {Boca
  Raton, FL},\ \bibinfo {year} {1985})\ pp.\ \bibinfo {pages}
  {65--137}\BibitemShut {NoStop}%
\bibitem [{\citenamefont {Natanson}\ \emph {et~al.}(1991)\citenamefont
  {Natanson}, \citenamefont {Garrett}, \citenamefont {Truong}, \citenamefont
  {Joseph},\ and\ \citenamefont {Truhlar}}]{truhlar91}%
  \BibitemOpen
  \bibfield  {author} {\bibinfo {author} {\bibfnamefont {G.~A.}\ \bibnamefont
  {Natanson}}, \bibinfo {author} {\bibfnamefont {B.~C.}\ \bibnamefont
  {Garrett}}, \bibinfo {author} {\bibfnamefont {T.~N.}\ \bibnamefont {Truong}},
  \bibinfo {author} {\bibfnamefont {T.}~\bibnamefont {Joseph}}, \ and\ \bibinfo
  {author} {\bibfnamefont {D.~G.}\ \bibnamefont {Truhlar}},\ }\href@noop {}
  {\bibfield  {journal} {\bibinfo  {journal} {J. Chem. Phys.}\ }\textbf
  {\bibinfo {volume} {94}},\ \bibinfo {pages} {7875} (\bibinfo {year}
  {1991})}\BibitemShut {NoStop}%
\bibitem [{\citenamefont {Truhlar}\ \emph {et~al.}(1996)\citenamefont
  {Truhlar}, \citenamefont {Garrett},\ and\ \citenamefont
  {Klippenstein}}]{truh96}%
  \BibitemOpen
  \bibfield  {author} {\bibinfo {author} {\bibfnamefont {D.~G.}\ \bibnamefont
  {Truhlar}}, \bibinfo {author} {\bibfnamefont {B.~C.}\ \bibnamefont
  {Garrett}}, \ and\ \bibinfo {author} {\bibfnamefont {S.~J.}\ \bibnamefont
  {Klippenstein}},\ }\href@noop {} {\bibfield  {journal} {\bibinfo  {journal}
  {J. Phys. Chem.}\ }\textbf {\bibinfo {volume} {100}},\ \bibinfo {pages}
  {12771} (\bibinfo {year} {1996})}\BibitemShut {NoStop}%
\bibitem [{\citenamefont {Truhlar}\ and\ \citenamefont
  {Garrett}(2000)}]{truh2000}%
  \BibitemOpen
  \bibfield  {author} {\bibinfo {author} {\bibfnamefont {D.~G.}\ \bibnamefont
  {Truhlar}}\ and\ \bibinfo {author} {\bibfnamefont {B.~C.}\ \bibnamefont
  {Garrett}},\ }\href@noop {} {\bibfield  {journal} {\bibinfo  {journal} {J.
  Phys. Chem. B}\ }\textbf {\bibinfo {volume} {104}},\ \bibinfo {pages} {1069}
  (\bibinfo {year} {2000})}\BibitemShut {NoStop}%
\bibitem [{\citenamefont {Komatsuzaki}\ and\ \citenamefont
  {Berry}(2001)}]{Komatsuzaki2001}%
  \BibitemOpen
  \bibfield  {author} {\bibinfo {author} {\bibfnamefont {T.}~\bibnamefont
  {Komatsuzaki}}\ and\ \bibinfo {author} {\bibfnamefont {R.~S.}\ \bibnamefont
  {Berry}},\ }\href@noop {} {\bibfield  {journal} {\bibinfo  {journal} {Proc.
  Natl. Acad. Sci. U.S.A.}\ }\textbf {\bibinfo {volume} {98}},\ \bibinfo
  {pages} {7666} (\bibinfo {year} {2001})}\BibitemShut {NoStop}%
\bibitem [{\citenamefont {Waalkens}\ \emph {et~al.}(2008)\citenamefont
  {Waalkens}, \citenamefont {Schubert},\ and\ \citenamefont
  {Wiggins}}]{Waalkens2008}%
  \BibitemOpen
  \bibfield  {author} {\bibinfo {author} {\bibfnamefont {H.}~\bibnamefont
  {Waalkens}}, \bibinfo {author} {\bibfnamefont {R.}~\bibnamefont {Schubert}},
  \ and\ \bibinfo {author} {\bibfnamefont {S.}~\bibnamefont {Wiggins}},\
  }\href@noop {} {\bibfield  {journal} {\bibinfo  {journal} {Nonlinearity}\
  }\textbf {\bibinfo {volume} {21}},\ \bibinfo {pages} {R1} (\bibinfo {year}
  {2008})}\BibitemShut {NoStop}%
\bibitem [{\citenamefont {Bartsch}\ \emph {et~al.}(2008)\citenamefont
  {Bartsch}, \citenamefont {Moix}, \citenamefont {Hernandez}, \citenamefont
  {Kawai},\ and\ \citenamefont {Uzer}}]{hern08d}%
  \BibitemOpen
  \bibfield  {author} {\bibinfo {author} {\bibfnamefont {T.}~\bibnamefont
  {Bartsch}}, \bibinfo {author} {\bibfnamefont {J.~M.}\ \bibnamefont {Moix}},
  \bibinfo {author} {\bibfnamefont {R.}~\bibnamefont {Hernandez}}, \bibinfo
  {author} {\bibfnamefont {S.}~\bibnamefont {Kawai}}, \ and\ \bibinfo {author}
  {\bibfnamefont {T.}~\bibnamefont {Uzer}},\ }\href@noop {} {\bibfield
  {journal} {\bibinfo  {journal} {Adv. Chem. Phys.}\ }\textbf {\bibinfo
  {volume} {140}},\ \bibinfo {pages} {191} (\bibinfo {year}
  {2008})}\BibitemShut {NoStop}%
\bibitem [{\citenamefont {Kawai}\ and\ \citenamefont
  {Komatsuzaki}(2010)}]{Komatsuzaki2010}%
  \BibitemOpen
  \bibfield  {author} {\bibinfo {author} {\bibfnamefont {S.}~\bibnamefont
  {Kawai}}\ and\ \bibinfo {author} {\bibfnamefont {T.}~\bibnamefont
  {Komatsuzaki}},\ }\href@noop {} {\bibfield  {journal} {\bibinfo  {journal}
  {Phys. Rev. Lett.}\ }\textbf {\bibinfo {volume} {105}},\ \bibinfo {pages}
  {048304} (\bibinfo {year} {2010})}\BibitemShut {NoStop}%
\bibitem [{\citenamefont {Hernandez}\ \emph {et~al.}(2010)\citenamefont
  {Hernandez}, \citenamefont {Bartsch},\ and\ \citenamefont {Uzer}}]{hern10a}%
  \BibitemOpen
  \bibfield  {author} {\bibinfo {author} {\bibfnamefont {R.}~\bibnamefont
  {Hernandez}}, \bibinfo {author} {\bibfnamefont {T.}~\bibnamefont {Bartsch}},
  \ and\ \bibinfo {author} {\bibfnamefont {T.}~\bibnamefont {Uzer}},\
  }\href@noop {} {\bibfield  {journal} {\bibinfo  {journal} {Chem. Phys.}\
  }\textbf {\bibinfo {volume} {370}},\ \bibinfo {pages} {270} (\bibinfo {year}
  {2010})}\BibitemShut {NoStop}%
\bibitem [{\citenamefont {Sharia}\ and\ \citenamefont
  {Henkelman}(2016)}]{Henkelman2016}%
  \BibitemOpen
  \bibfield  {author} {\bibinfo {author} {\bibfnamefont {O.}~\bibnamefont
  {Sharia}}\ and\ \bibinfo {author} {\bibfnamefont {G.}~\bibnamefont
  {Henkelman}},\ }\href@noop {} {\bibfield  {journal} {\bibinfo  {journal} {New
  J. Phys.}\ }\textbf {\bibinfo {volume} {18}},\ \bibinfo {pages} {013023}
  (\bibinfo {year} {2016})}\BibitemShut {NoStop}%
\bibitem [{\citenamefont {Lichtenberg}\ and\ \citenamefont
  {Liebermann}(1982)}]{Lichtenberg82}%
  \BibitemOpen
  \bibfield  {author} {\bibinfo {author} {\bibfnamefont {A.~J.}\ \bibnamefont
  {Lichtenberg}}\ and\ \bibinfo {author} {\bibfnamefont {M.~A.}\ \bibnamefont
  {Liebermann}},\ }\href@noop {} {\emph {\bibinfo {title} {Regular and
  Stochastic Motion}}}\ (\bibinfo  {publisher} {Springer},\ \bibinfo {address}
  {New York},\ \bibinfo {year} {1982})\BibitemShut {NoStop}%
\bibitem [{\citenamefont {Hernandez}\ \emph {et~al.}(1993)\citenamefont
  {Hernandez}, \citenamefont {Miller}, \citenamefont {Moore},\ and\
  \citenamefont {Polik}}]{hern93a}%
  \BibitemOpen
  \bibfield  {author} {\bibinfo {author} {\bibfnamefont {R.}~\bibnamefont
  {Hernandez}}, \bibinfo {author} {\bibfnamefont {W.~H.}\ \bibnamefont
  {Miller}}, \bibinfo {author} {\bibfnamefont {C.~B.}\ \bibnamefont {Moore}}, \
  and\ \bibinfo {author} {\bibfnamefont {W.~F.}\ \bibnamefont {Polik}},\
  }\href@noop {} {\bibfield  {journal} {\bibinfo  {journal} {J. Chem. Phys.}\
  }\textbf {\bibinfo {volume} {99}},\ \bibinfo {pages} {950} (\bibinfo {year}
  {1993})}\BibitemShut {NoStop}%
\bibitem [{\citenamefont {Ott}(2002)}]{Ott2002a}%
  \BibitemOpen
  \bibfield  {author} {\bibinfo {author} {\bibfnamefont {E.}~\bibnamefont
  {Ott}},\ }\href@noop {} {\emph {\bibinfo {title} {Chaos in dynamical
  systems}}},\ \bibinfo {edition} {second edition}\ ed.\ (\bibinfo  {publisher}
  {Cambridge University Press},\ \bibinfo {address} {Cambridge},\ \bibinfo
  {year} {2002})\BibitemShut {NoStop}%
\bibitem [{\citenamefont {Wiggins}(2013)}]{wiggins2013normally}%
  \BibitemOpen
  \bibfield  {author} {\bibinfo {author} {\bibfnamefont {S.}~\bibnamefont
  {Wiggins}},\ }\href@noop {} {\emph {\bibinfo {title} {Normally hyperbolic
  invariant manifolds in dynamical systems}}},\ Vol.\ \bibinfo {volume} {105}\
  (\bibinfo  {publisher} {Springer Science \& Business Media},\ \bibinfo {year}
  {2013})\BibitemShut {NoStop}%
\bibitem [{\citenamefont {Pollak}\ and\ \citenamefont
  {Pechukas}(1978)}]{pollak78}%
  \BibitemOpen
  \bibfield  {author} {\bibinfo {author} {\bibfnamefont {E.}~\bibnamefont
  {Pollak}}\ and\ \bibinfo {author} {\bibfnamefont {P.}~\bibnamefont
  {Pechukas}},\ }\href@noop {} {\bibfield  {journal} {\bibinfo  {journal} {J.
  Chem. Phys.}\ }\textbf {\bibinfo {volume} {69}},\ \bibinfo {pages} {1218}
  (\bibinfo {year} {1978})}\BibitemShut {NoStop}%
\bibitem [{\citenamefont {Pechukas}\ and\ \citenamefont
  {Pollak}(1979)}]{pech79a}%
  \BibitemOpen
  \bibfield  {author} {\bibinfo {author} {\bibfnamefont {P.}~\bibnamefont
  {Pechukas}}\ and\ \bibinfo {author} {\bibfnamefont {E.}~\bibnamefont
  {Pollak}},\ }\href@noop {} {\bibfield  {journal} {\bibinfo  {journal} {J.
  Chem. Phys.}\ }\textbf {\bibinfo {volume} {71}},\ \bibinfo {pages} {2062}
  (\bibinfo {year} {1979})}\BibitemShut {NoStop}%
\bibitem [{\citenamefont {Hernandez}\ and\ \citenamefont
  {Miller}(1993)}]{hern93b}%
  \BibitemOpen
  \bibfield  {author} {\bibinfo {author} {\bibfnamefont {R.}~\bibnamefont
  {Hernandez}}\ and\ \bibinfo {author} {\bibfnamefont {W.~H.}\ \bibnamefont
  {Miller}},\ }\href@noop {} {\bibfield  {journal} {\bibinfo  {journal} {Chem.
  Phys. Lett.}\ }\textbf {\bibinfo {volume} {214}},\ \bibinfo {pages} {129}
  (\bibinfo {year} {1993})}\BibitemShut {NoStop}%
\bibitem [{\citenamefont {Hernandez}(1994)}]{hern94}%
  \BibitemOpen
  \bibfield  {author} {\bibinfo {author} {\bibfnamefont {R.}~\bibnamefont
  {Hernandez}},\ }\href@noop {} {\bibfield  {journal} {\bibinfo  {journal} {J.
  Chem. Phys.}\ }\textbf {\bibinfo {volume} {101}},\ \bibinfo {pages} {9534}
  (\bibinfo {year} {1994})}\BibitemShut {NoStop}%
\bibitem [{\citenamefont {Uzer}\ \emph {et~al.}(2002)\citenamefont {Uzer},
  \citenamefont {Jaff{\'e}}, \citenamefont {Palaci{\'a}n}, \citenamefont
  {Yanguas},\ and\ \citenamefont {Wiggins}}]{Uzer02}%
  \BibitemOpen
  \bibfield  {author} {\bibinfo {author} {\bibfnamefont {T.}~\bibnamefont
  {Uzer}}, \bibinfo {author} {\bibfnamefont {C.}~\bibnamefont {Jaff{\'e}}},
  \bibinfo {author} {\bibfnamefont {J.}~\bibnamefont {Palaci{\'a}n}}, \bibinfo
  {author} {\bibfnamefont {P.}~\bibnamefont {Yanguas}}, \ and\ \bibinfo
  {author} {\bibfnamefont {S.}~\bibnamefont {Wiggins}},\ }\href@noop {}
  {\bibfield  {journal} {\bibinfo  {journal} {Nonlinearity}\ }\textbf {\bibinfo
  {volume} {15}},\ \bibinfo {pages} {957} (\bibinfo {year} {2002})}\BibitemShut
  {NoStop}%
\bibitem [{\citenamefont {Teramoto}\ \emph {et~al.}(2011)\citenamefont
  {Teramoto}, \citenamefont {Toda},\ and\ \citenamefont
  {Komatsuzaki}}]{Teramoto11}%
  \BibitemOpen
  \bibfield  {author} {\bibinfo {author} {\bibfnamefont {H.}~\bibnamefont
  {Teramoto}}, \bibinfo {author} {\bibfnamefont {M.}~\bibnamefont {Toda}}, \
  and\ \bibinfo {author} {\bibfnamefont {T.}~\bibnamefont {Komatsuzaki}},\
  }\href@noop {} {\bibfield  {journal} {\bibinfo  {journal} {Phys. Rev. Lett.}\
  }\textbf {\bibinfo {volume} {106}},\ \bibinfo {pages} {054101} (\bibinfo
  {year} {2011})}\BibitemShut {NoStop}%
\bibitem [{\citenamefont {Li}\ \emph {et~al.}(2006)\citenamefont {Li},
  \citenamefont {Shoujiguchi}, \citenamefont {Toda},\ and\ \citenamefont
  {Komatsuzaki}}]{Li06prl}%
  \BibitemOpen
  \bibfield  {author} {\bibinfo {author} {\bibfnamefont {C.-B.}\ \bibnamefont
  {Li}}, \bibinfo {author} {\bibfnamefont {A.}~\bibnamefont {Shoujiguchi}},
  \bibinfo {author} {\bibfnamefont {M.}~\bibnamefont {Toda}}, \ and\ \bibinfo
  {author} {\bibfnamefont {T.}~\bibnamefont {Komatsuzaki}},\ }\href@noop {}
  {\bibfield  {journal} {\bibinfo  {journal} {Phys. Rev. Lett.}\ }\textbf
  {\bibinfo {volume} {97}},\ \bibinfo {pages} {028302} (\bibinfo {year}
  {2006})}\BibitemShut {NoStop}%
\bibitem [{\citenamefont {Waalkens}\ and\ \citenamefont
  {Wiggins}(2004)}]{Waalkens04b}%
  \BibitemOpen
  \bibfield  {author} {\bibinfo {author} {\bibfnamefont {H.}~\bibnamefont
  {Waalkens}}\ and\ \bibinfo {author} {\bibfnamefont {S.}~\bibnamefont
  {Wiggins}},\ }\href {\doibase 10.1088/0305-4470/37/35/L02} {\bibfield
  {journal} {\bibinfo  {journal} {J. Phys. A}\ }\textbf {\bibinfo {volume}
  {37}},\ \bibinfo {pages} {L435} (\bibinfo {year} {2004})}\BibitemShut
  {NoStop}%
\bibitem [{\citenamefont {\ifmmode \mbox{\c{C}}\else
  \c{C}\fi{}ift\ifmmode~\mbox{\c{c}}\else \c{c}\fi{}i}\ and\ \citenamefont
  {Waalkens}(2013)}]{Waalkens13}%
  \BibitemOpen
  \bibfield  {author} {\bibinfo {author} {\bibfnamefont {U.}~\bibnamefont
  {\ifmmode \mbox{\c{C}}\else \c{C}\fi{}ift\ifmmode~\mbox{\c{c}}\else
  \c{c}\fi{}i}}\ and\ \bibinfo {author} {\bibfnamefont {H.}~\bibnamefont
  {Waalkens}},\ }\href@noop {} {\bibfield  {journal} {\bibinfo  {journal}
  {Phys. Rev. Lett.}\ }\textbf {\bibinfo {volume} {110}},\ \bibinfo {pages}
  {233201} (\bibinfo {year} {2013})}\BibitemShut {NoStop}%
\bibitem [{\citenamefont {Mendoza}\ and\ \citenamefont
  {Mancho}(2010)}]{Mancho2010}%
  \BibitemOpen
  \bibfield  {author} {\bibinfo {author} {\bibfnamefont {C.}~\bibnamefont
  {Mendoza}}\ and\ \bibinfo {author} {\bibfnamefont {A.~M.}\ \bibnamefont
  {Mancho}},\ }\href@noop {} {\bibfield  {journal} {\bibinfo  {journal} {Phys.
  Rev. Lett.}\ }\textbf {\bibinfo {volume} {105}},\ \bibinfo {pages} {038501}
  (\bibinfo {year} {2010})}\BibitemShut {NoStop}%
\bibitem [{\citenamefont {Mancho}\ \emph {et~al.}(2013)\citenamefont {Mancho},
  \citenamefont {Wiggins}, \citenamefont {Curbelo},\ and\ \citenamefont
  {Mendoza}}]{Mancho2013}%
  \BibitemOpen
  \bibfield  {author} {\bibinfo {author} {\bibfnamefont {A.~M.}\ \bibnamefont
  {Mancho}}, \bibinfo {author} {\bibfnamefont {S.}~\bibnamefont {Wiggins}},
  \bibinfo {author} {\bibfnamefont {J.}~\bibnamefont {Curbelo}}, \ and\
  \bibinfo {author} {\bibfnamefont {C.}~\bibnamefont {Mendoza}},\ }\href@noop
  {} {\bibfield  {journal} {\bibinfo  {journal} {Commun. Nonlinear Sci. Numer.
  Simul.}\ }\textbf {\bibinfo {volume} {18}},\ \bibinfo {pages} {3530 }
  (\bibinfo {year} {2013})}\BibitemShut {NoStop}%
\bibitem [{\citenamefont {Craven}\ and\ \citenamefont
  {Hernandez}(2015)}]{hern15e}%
  \BibitemOpen
  \bibfield  {author} {\bibinfo {author} {\bibfnamefont {G.~T.}\ \bibnamefont
  {Craven}}\ and\ \bibinfo {author} {\bibfnamefont {R.}~\bibnamefont
  {Hernandez}},\ }\href@noop {} {\bibfield  {journal} {\bibinfo  {journal}
  {Phys. Rev. Lett.}\ }\textbf {\bibinfo {volume} {115}},\ \bibinfo {pages}
  {148301} (\bibinfo {year} {2015})}\BibitemShut {NoStop}%
\bibitem [{\citenamefont {Craven}\ and\ \citenamefont
  {Hernandez}(2016)}]{hern16d}%
  \BibitemOpen
  \bibfield  {author} {\bibinfo {author} {\bibfnamefont {G.~T.}\ \bibnamefont
  {Craven}}\ and\ \bibinfo {author} {\bibfnamefont {R.}~\bibnamefont
  {Hernandez}},\ }\href@noop {} {\bibfield  {journal} {\bibinfo  {journal}
  {Phys. Chem. Chem. Phys.}\ }\textbf {\bibinfo {volume} {18}},\ \bibinfo
  {pages} {4008} (\bibinfo {year} {2016})}\BibitemShut {NoStop}%
\bibitem [{\citenamefont {Junginger}\ \emph {et~al.}(2016)\citenamefont
  {Junginger}, \citenamefont {Craven}, \citenamefont {Bartsch}, \citenamefont
  {Revuelta}, \citenamefont {Borondo}, \citenamefont {Benito},\ and\
  \citenamefont {Hernandez}}]{hern16h}%
  \BibitemOpen
  \bibfield  {author} {\bibinfo {author} {\bibfnamefont {A.}~\bibnamefont
  {Junginger}}, \bibinfo {author} {\bibfnamefont {G.~T.}\ \bibnamefont
  {Craven}}, \bibinfo {author} {\bibfnamefont {T.}~\bibnamefont {Bartsch}},
  \bibinfo {author} {\bibfnamefont {F.}~\bibnamefont {Revuelta}}, \bibinfo
  {author} {\bibfnamefont {F.}~\bibnamefont {Borondo}}, \bibinfo {author}
  {\bibfnamefont {R.~M.}\ \bibnamefont {Benito}}, \ and\ \bibinfo {author}
  {\bibfnamefont {R.}~\bibnamefont {Hernandez}},\ }\href@noop {} {\bibfield
  {journal} {\bibinfo  {journal} {Phys. Chem. Chem. Phys.}\ }\textbf {\bibinfo
  {volume} {18}},\ \bibinfo {pages} {30270} (\bibinfo {year}
  {2016})}\BibitemShut {NoStop}%
\bibitem [{\citenamefont {Junginger}\ and\ \citenamefont
  {Hernandez}(2016{\natexlab{a}})}]{hern16i}%
  \BibitemOpen
  \bibfield  {author} {\bibinfo {author} {\bibfnamefont {A.}~\bibnamefont
  {Junginger}}\ and\ \bibinfo {author} {\bibfnamefont {R.}~\bibnamefont
  {Hernandez}},\ }\href@noop {} {\bibfield  {journal} {\bibinfo  {journal}
  {Phys. Chem. Chem. Phys.}\ }\textbf {\bibinfo {volume} {18}},\ \bibinfo
  {pages} {30282} (\bibinfo {year} {2016}{\natexlab{a}})}\BibitemShut {NoStop}%
\bibitem [{\citenamefont {Feldmaier}\ \emph {et~al.}(2017)\citenamefont
  {Feldmaier}, \citenamefont {Junginger}, \citenamefont {Main}, \citenamefont
  {Wunner},\ and\ \citenamefont {Hernandez}}]{hern17h}%
  \BibitemOpen
  \bibfield  {author} {\bibinfo {author} {\bibfnamefont {M.}~\bibnamefont
  {Feldmaier}}, \bibinfo {author} {\bibfnamefont {A.}~\bibnamefont
  {Junginger}}, \bibinfo {author} {\bibfnamefont {J.}~\bibnamefont {Main}},
  \bibinfo {author} {\bibfnamefont {G.}~\bibnamefont {Wunner}}, \ and\ \bibinfo
  {author} {\bibfnamefont {R.}~\bibnamefont {Hernandez}},\ }\href@noop {}
  {\bibfield  {journal} {\bibinfo  {journal} {Chem. Phys. Lett.}\ }\textbf
  {\bibinfo {volume} {687}},\ \bibinfo {pages} {194} (\bibinfo {year}
  {2017})}\BibitemShut {NoStop}%
\bibitem [{\citenamefont {Schraft}\ \emph {et~al.}(2018)\citenamefont
  {Schraft}, \citenamefont {Junginger}, \citenamefont {Feldmaier},
  \citenamefont {Bardakcioglu}, \citenamefont {Main}, \citenamefont {Wunner},\
  and\ \citenamefont {Hernandez}}]{hern18c}%
  \BibitemOpen
  \bibfield  {author} {\bibinfo {author} {\bibfnamefont {P.}~\bibnamefont
  {Schraft}}, \bibinfo {author} {\bibfnamefont {A.}~\bibnamefont {Junginger}},
  \bibinfo {author} {\bibfnamefont {M.}~\bibnamefont {Feldmaier}}, \bibinfo
  {author} {\bibfnamefont {R.}~\bibnamefont {Bardakcioglu}}, \bibinfo {author}
  {\bibfnamefont {J.}~\bibnamefont {Main}}, \bibinfo {author} {\bibfnamefont
  {G.}~\bibnamefont {Wunner}}, \ and\ \bibinfo {author} {\bibfnamefont
  {R.}~\bibnamefont {Hernandez}},\ }\href@noop {} {\bibfield  {journal}
  {\bibinfo  {journal} {Phys. Rev. E}\ }\textbf {\bibinfo {volume} {97}},\
  \bibinfo {pages} {042309} (\bibinfo {year} {2018})}\BibitemShut {NoStop}%
\bibitem [{\citenamefont {Junginger}\ and\ \citenamefont
  {Hernandez}(2016{\natexlab{b}})}]{hern16a}%
  \BibitemOpen
  \bibfield  {author} {\bibinfo {author} {\bibfnamefont {A.}~\bibnamefont
  {Junginger}}\ and\ \bibinfo {author} {\bibfnamefont {R.}~\bibnamefont
  {Hernandez}},\ }\href@noop {} {\bibfield  {journal} {\bibinfo  {journal} {J.
  Phys. Chem. B}\ }\textbf {\bibinfo {volume} {120}},\ \bibinfo {pages} {1720}
  (\bibinfo {year} {2016}{\natexlab{b}})}\BibitemShut {NoStop}%
\bibitem [{\citenamefont {Bartsch}\ \emph
  {et~al.}(2005{\natexlab{a}})\citenamefont {Bartsch}, \citenamefont
  {Hernandez},\ and\ \citenamefont {Uzer}}]{dawn05a}%
  \BibitemOpen
  \bibfield  {author} {\bibinfo {author} {\bibfnamefont {T.}~\bibnamefont
  {Bartsch}}, \bibinfo {author} {\bibfnamefont {R.}~\bibnamefont {Hernandez}},
  \ and\ \bibinfo {author} {\bibfnamefont {T.}~\bibnamefont {Uzer}},\
  }\href@noop {} {\bibfield  {journal} {\bibinfo  {journal} {Phys. Rev. Lett.}\
  }\textbf {\bibinfo {volume} {95}},\ \bibinfo {pages} {058301} (\bibinfo
  {year} {2005}{\natexlab{a}})}\BibitemShut {NoStop}%
\bibitem [{\citenamefont {Bartsch}\ \emph
  {et~al.}(2005{\natexlab{b}})\citenamefont {Bartsch}, \citenamefont {Uzer},\
  and\ \citenamefont {Hernandez}}]{dawn05b}%
  \BibitemOpen
  \bibfield  {author} {\bibinfo {author} {\bibfnamefont {T.}~\bibnamefont
  {Bartsch}}, \bibinfo {author} {\bibfnamefont {T.}~\bibnamefont {Uzer}}, \
  and\ \bibinfo {author} {\bibfnamefont {R.}~\bibnamefont {Hernandez}},\
  }\href@noop {} {\bibfield  {journal} {\bibinfo  {journal} {J. Chem. Phys.}\
  }\textbf {\bibinfo {volume} {123}},\ \bibinfo {pages} {204102} (\bibinfo
  {year} {2005}{\natexlab{b}})}\BibitemShut {NoStop}%
\bibitem [{\citenamefont {Bartsch}\ \emph {et~al.}(2006)\citenamefont
  {Bartsch}, \citenamefont {Uzer}, \citenamefont {Moix},\ and\ \citenamefont
  {Hernandez}}]{hern06d}%
  \BibitemOpen
  \bibfield  {author} {\bibinfo {author} {\bibfnamefont {T.}~\bibnamefont
  {Bartsch}}, \bibinfo {author} {\bibfnamefont {T.}~\bibnamefont {Uzer}},
  \bibinfo {author} {\bibfnamefont {J.~M.}\ \bibnamefont {Moix}}, \ and\
  \bibinfo {author} {\bibfnamefont {R.}~\bibnamefont {Hernandez}},\ }\href@noop
  {} {\bibfield  {journal} {\bibinfo  {journal} {J. Chem. Phys.}\ }\textbf
  {\bibinfo {volume} {124}},\ \bibinfo {pages} {244310} (\bibinfo {year}
  {2006})}\BibitemShut {NoStop}%
\bibitem [{\citenamefont {Haller}\ \emph {et~al.}(2010)\citenamefont {Haller},
  \citenamefont {Uzer}, \citenamefont {Palaci{\'a}n}, \citenamefont {Yanguas},\
  and\ \citenamefont {Jaff{\'e}}}]{jaffe10}%
  \BibitemOpen
  \bibfield  {author} {\bibinfo {author} {\bibfnamefont {G.}~\bibnamefont
  {Haller}}, \bibinfo {author} {\bibfnamefont {T.}~\bibnamefont {Uzer}},
  \bibinfo {author} {\bibfnamefont {J.}~\bibnamefont {Palaci{\'a}n}}, \bibinfo
  {author} {\bibfnamefont {P.}~\bibnamefont {Yanguas}}, \ and\ \bibinfo
  {author} {\bibfnamefont {C.}~\bibnamefont {Jaff{\'e}}},\ }\href@noop {}
  {\bibfield  {journal} {\bibinfo  {journal} {Communications in Nonlinear
  Science and Numerical Simulation}\ }\textbf {\bibinfo {volume} {15}},\
  \bibinfo {pages} {48} (\bibinfo {year} {2010})}\BibitemShut {NoStop}%
\bibitem [{\citenamefont {Haller}\ \emph {et~al.}(2011)\citenamefont {Haller},
  \citenamefont {Uzer}, \citenamefont {Palaci\'{a}n}, \citenamefont {Yanguas},\
  and\ \citenamefont {Jaff{\'e}}}]{jaffe11}%
  \BibitemOpen
  \bibfield  {author} {\bibinfo {author} {\bibfnamefont {G.}~\bibnamefont
  {Haller}}, \bibinfo {author} {\bibfnamefont {T.}~\bibnamefont {Uzer}},
  \bibinfo {author} {\bibfnamefont {J.}~\bibnamefont {Palaci\'{a}n}}, \bibinfo
  {author} {\bibfnamefont {P.}~\bibnamefont {Yanguas}}, \ and\ \bibinfo
  {author} {\bibfnamefont {C.}~\bibnamefont {Jaff{\'e}}},\ }\href@noop {}
  {\bibfield  {journal} {\bibinfo  {journal} {Nonlinearity}\ }\textbf {\bibinfo
  {volume} {24}},\ \bibinfo {pages} {527} (\bibinfo {year} {2011})}\BibitemShut
  {NoStop}%
\bibitem [{\citenamefont {Nagahata}\ \emph {et~al.}(2013)\citenamefont
  {Nagahata}, \citenamefont {Teramoto}, \citenamefont {Li}, \citenamefont
  {Kawai},\ and\ \citenamefont {Komatsuzaki}}]{komatsuzaki13a}%
  \BibitemOpen
  \bibfield  {author} {\bibinfo {author} {\bibfnamefont {Y.}~\bibnamefont
  {Nagahata}}, \bibinfo {author} {\bibfnamefont {H.}~\bibnamefont {Teramoto}},
  \bibinfo {author} {\bibfnamefont {C.-B.}\ \bibnamefont {Li}}, \bibinfo
  {author} {\bibfnamefont {S.}~\bibnamefont {Kawai}}, \ and\ \bibinfo {author}
  {\bibfnamefont {T.}~\bibnamefont {Komatsuzaki}},\ }\href@noop {} {\bibfield
  {journal} {\bibinfo  {journal} {Phys. Rev. E}\ }\textbf {\bibinfo {volume}
  {88}},\ \bibinfo {pages} {042923} (\bibinfo {year} {2013})}\BibitemShut
  {NoStop}%
\bibitem [{\citenamefont {Townsend}\ \emph {et~al.}(2004)\citenamefont
  {Townsend}, \citenamefont {Lahankar}, \citenamefont {Lee}, \citenamefont
  {Chambreau}, \citenamefont {Suits}, \citenamefont {Zhang}, \citenamefont
  {Rheinecker}, \citenamefont {Harding},\ and\ \citenamefont
  {Bowman}}]{bowman04a}%
  \BibitemOpen
  \bibfield  {author} {\bibinfo {author} {\bibfnamefont {D.}~\bibnamefont
  {Townsend}}, \bibinfo {author} {\bibfnamefont {S.~A.}\ \bibnamefont
  {Lahankar}}, \bibinfo {author} {\bibfnamefont {S.~K.}\ \bibnamefont {Lee}},
  \bibinfo {author} {\bibfnamefont {S.~D.}\ \bibnamefont {Chambreau}}, \bibinfo
  {author} {\bibfnamefont {A.~G.}\ \bibnamefont {Suits}}, \bibinfo {author}
  {\bibfnamefont {X.}~\bibnamefont {Zhang}}, \bibinfo {author} {\bibfnamefont
  {J.~L.}\ \bibnamefont {Rheinecker}}, \bibinfo {author} {\bibfnamefont
  {L.~B.}\ \bibnamefont {Harding}}, \ and\ \bibinfo {author} {\bibfnamefont
  {J.~M.}\ \bibnamefont {Bowman}},\ }\href@noop {} {\bibfield  {journal}
  {\bibinfo  {journal} {Science}\ }\textbf {\bibinfo {volume} {306}},\ \bibinfo
  {pages} {1158} (\bibinfo {year} {2004})}\BibitemShut {NoStop}%
\bibitem [{\citenamefont {Ulusoy}\ \emph {et~al.}(2013)\citenamefont {Ulusoy},
  \citenamefont {Stanton},\ and\ \citenamefont {Hernandez}}]{hern13e}%
  \BibitemOpen
  \bibfield  {author} {\bibinfo {author} {\bibfnamefont {I.~S.}\ \bibnamefont
  {Ulusoy}}, \bibinfo {author} {\bibfnamefont {J.~F.}\ \bibnamefont {Stanton}},
  \ and\ \bibinfo {author} {\bibfnamefont {R.}~\bibnamefont {Hernandez}},\
  }\href@noop {} {\bibfield  {journal} {\bibinfo  {journal} {J. Phys. Chem. A}\
  }\textbf {\bibinfo {volume} {117}},\ \bibinfo {pages} {10567} (\bibinfo
  {year} {2013})}\BibitemShut {NoStop}%
\bibitem [{\citenamefont {Maugi\`{e}re}\ \emph {et~al.}(2014)\citenamefont
  {Maugi\`{e}re}, \citenamefont {Collins}, \citenamefont {Ezra}, \citenamefont
  {Farantos},\ and\ \citenamefont {Wiggins}}]{wiggins14a}%
  \BibitemOpen
  \bibfield  {author} {\bibinfo {author} {\bibfnamefont {F.~A.~L.}\
  \bibnamefont {Maugi\`{e}re}}, \bibinfo {author} {\bibfnamefont
  {P.}~\bibnamefont {Collins}}, \bibinfo {author} {\bibfnamefont
  {G.}~\bibnamefont {Ezra}}, \bibinfo {author} {\bibfnamefont {S.~C.}\
  \bibnamefont {Farantos}}, \ and\ \bibinfo {author} {\bibfnamefont
  {S.}~\bibnamefont {Wiggins}},\ }\href@noop {} {\bibfield  {journal} {\bibinfo
   {journal} {Chem. Phys. Lett.}\ }\textbf {\bibinfo {volume} {592}},\ \bibinfo
  {pages} {282} (\bibinfo {year} {2014})}\BibitemShut {NoStop}%
\bibitem [{\citenamefont {Bowman}\ and\ \citenamefont
  {Suits}(2011)}]{bowman2011c}%
  \BibitemOpen
  \bibfield  {author} {\bibinfo {author} {\bibfnamefont {J.~M.}\ \bibnamefont
  {Bowman}}\ and\ \bibinfo {author} {\bibfnamefont {A.~G.}\ \bibnamefont
  {Suits}},\ }\href@noop {} {\bibfield  {journal} {\bibinfo  {journal} {Phys.
  Today}\ }\textbf {\bibinfo {volume} {64}},\ \bibinfo {pages} {33} (\bibinfo
  {year} {2011})}\BibitemShut {NoStop}%
\bibitem [{\citenamefont {Bowman}(2014)}]{bowman14a}%
  \BibitemOpen
  \bibfield  {author} {\bibinfo {author} {\bibfnamefont {J.~M.}\ \bibnamefont
  {Bowman}},\ }\href@noop {} {\bibfield  {journal} {\bibinfo  {journal} {Mol.
  Phys.}\ }\textbf {\bibinfo {volume} {112}},\ \bibinfo {pages} {2516}
  (\bibinfo {year} {2014})}\BibitemShut {NoStop}%
\bibitem [{\citenamefont {Miller}(1974)}]{mill74}%
  \BibitemOpen
  \bibfield  {author} {\bibinfo {author} {\bibfnamefont {W.~H.}\ \bibnamefont
  {Miller}},\ }\href@noop {} {\bibfield  {journal} {\bibinfo  {journal} {J.
  Chem. Phys.}\ }\textbf {\bibinfo {volume} {61}},\ \bibinfo {pages} {1823}
  (\bibinfo {year} {1974})}\BibitemShut {NoStop}%
\bibitem [{\citenamefont {Garrett}\ \emph {et~al.}(1982)\citenamefont
  {Garrett}, \citenamefont {Isaacson}, \citenamefont {Skodje},\ and\
  \citenamefont {Truhlar}}]{truh82}%
  \BibitemOpen
  \bibfield  {author} {\bibinfo {author} {\bibfnamefont {B.~C.}\ \bibnamefont
  {Garrett}}, \bibinfo {author} {\bibfnamefont {A.~D.}\ \bibnamefont
  {Isaacson}}, \bibinfo {author} {\bibfnamefont {R.~T.}\ \bibnamefont
  {Skodje}}, \ and\ \bibinfo {author} {\bibfnamefont {D.~G.}\ \bibnamefont
  {Truhlar}},\ }\href@noop {} {\bibfield  {journal} {\bibinfo  {journal} {J.
  Phys. Chem.}\ }\textbf {\bibinfo {volume} {86}},\ \bibinfo {pages} {2252}
  (\bibinfo {year} {1982})}\BibitemShut {NoStop}%
\bibitem [{\citenamefont {Miller}(1977)}]{mill77}%
  \BibitemOpen
  \bibfield  {author} {\bibinfo {author} {\bibfnamefont {W.~H.}\ \bibnamefont
  {Miller}},\ }\href@noop {} {\bibfield  {journal} {\bibinfo  {journal}
  {Faraday Discuss. Chem. Soc.}\ }\textbf {\bibinfo {volume} {62}},\ \bibinfo
  {pages} {40} (\bibinfo {year} {1977})}\BibitemShut {NoStop}%
\bibitem [{\citenamefont {Miller}\ \emph {et~al.}(1990)\citenamefont {Miller},
  \citenamefont {Hernandez}, \citenamefont {Handy}, \citenamefont
  {Jayatilaka},\ and\ \citenamefont {Willetts}}]{mill90b}%
  \BibitemOpen
  \bibfield  {author} {\bibinfo {author} {\bibfnamefont {W.~H.}\ \bibnamefont
  {Miller}}, \bibinfo {author} {\bibfnamefont {R.}~\bibnamefont {Hernandez}},
  \bibinfo {author} {\bibfnamefont {N.~C.}\ \bibnamefont {Handy}}, \bibinfo
  {author} {\bibfnamefont {D.}~\bibnamefont {Jayatilaka}}, \ and\ \bibinfo
  {author} {\bibfnamefont {A.}~\bibnamefont {Willetts}},\ }\href@noop {}
  {\bibfield  {journal} {\bibinfo  {journal} {Chem. Phys. Lett.}\ }\textbf
  {\bibinfo {volume} {172}},\ \bibinfo {pages} {62} (\bibinfo {year}
  {1990})}\BibitemShut {NoStop}%
\bibitem [{\citenamefont {Nguyen}\ \emph {et~al.}(2011)\citenamefont {Nguyen},
  \citenamefont {Stanton},\ and\ \citenamefont {Barker}}]{Stanton2011}%
  \BibitemOpen
  \bibfield  {author} {\bibinfo {author} {\bibfnamefont {T.~L.}\ \bibnamefont
  {Nguyen}}, \bibinfo {author} {\bibfnamefont {J.~F.}\ \bibnamefont {Stanton}},
  \ and\ \bibinfo {author} {\bibfnamefont {J.~R.}\ \bibnamefont {Barker}},\
  }\href@noop {} {\bibfield  {journal} {\bibinfo  {journal} {J. Phys. Chem. A}\
  }\textbf {\bibinfo {volume} {115}},\ \bibinfo {pages} {5118} (\bibinfo {year}
  {2011})}\BibitemShut {NoStop}%
\bibitem [{\citenamefont {Barker}\ \emph {et~al.}(2012)\citenamefont {Barker},
  \citenamefont {Nguyen},\ and\ \citenamefont {Stanton}}]{Stanton2012}%
  \BibitemOpen
  \bibfield  {author} {\bibinfo {author} {\bibfnamefont {J.~R.}\ \bibnamefont
  {Barker}}, \bibinfo {author} {\bibfnamefont {T.~L.}\ \bibnamefont {Nguyen}},
  \ and\ \bibinfo {author} {\bibfnamefont {J.~F.}\ \bibnamefont {Stanton}},\
  }\href@noop {} {\bibfield  {journal} {\bibinfo  {journal} {J. Phys. Chem. A}\
  }\textbf {\bibinfo {volume} {116}},\ \bibinfo {pages} {6408} (\bibinfo {year}
  {2012})}\BibitemShut {NoStop}%
\end{thebibliography}%

\end{document}